\documentclass[aps,pre,10pt,twocolumn,superscriptaddress,nofootinbib]{revtex4-2}

\usepackage{amsmath,amssymb,times,hyperref,graphicx}
\hypersetup{colorlinks=true, linkcolor=blue, citecolor=blue, filecolor=blue, urlcolor=blue}

\usepackage{tabularray}
\usepackage{upgreek}
\usepackage{xcolor}

\renewcommand{\a}{\mathrm{a}}
\newcommand{\s}{\mathrm{s}}
\newcommand{\p}{\mathrm{p}}
\newcommand{\ns}{\mathrm{ns}}
\renewcommand{\d}{\mathrm{d}}
\newcommand{\kt}{k_\mathrm{B}T}
\newcommand{\micron}{\upmu\mathrm{m}}
\renewcommand{\max}{\mathrm{max}}
\newcommand{\spn}{\mathrm{sp}}
\newcommand{\bin}{\mathrm{bin}}
\newcommand{\tot}{\mathrm{tot}}
\newcommand{\mol}{\mathrm{mol}}
\newcommand{\col}{\mathrm{col}}
\newcommand{\ext}{\mathrm{ext}}

\newcommand{\eg}{\textit{e.g.}, }
\newcommand{\ie}{\textit{i.e.}, }
\newcommand{\cf}{\textit{cf.} }
\newcommand{\unit}[1]{\hat{\mathbf{#1}}}

\newcommand{\vecF}{\mathbf{F}}
\newcommand{\vecJ}{\mathbf{J}}
\newcommand{\vecq}{\mathbf{q}}
\newcommand{\vecr}{\mathbf{r}}
\newcommand{\vecv}{\mathbf{v}}
\newcommand{\grad}{\boldsymbol{\nabla}}

\begin{document} 

\title{Propelling Catalytic Structures Using Active Biomolecular Condensates} 

\author{Benjamin Sorkin}
\email{bs4171@princeton.edu}
\affiliation{Princeton Center for Theoretical Science, Princeton University, Princeton, New Jersey 08544, USA}

\author{Ned S. Wingreen}
\email{wingreen@princeton.edu}
\affiliation{Princeton Center for Theoretical Science, Princeton University, Princeton, New Jersey 08544, USA}
\affiliation{Lewis-Sigler Institute for Integrative Genomics, Princeton University, Princeton, New Jersey 08544, USA}
\affiliation{Department of Molecular Biology, Princeton University, Princeton, New Jersey 08544, USA}

\begin{abstract}
    Living systems routinely consume energy to achieve motility, often using intricate biomolecular machinery. In this work, we demonstrate that active biomolecular condensates enable the self-propulsion of micron-sized objects in otherwise homogeneous, isotropic, and autonomous environments. Our proposed minimal mechanism consists of phase-separating proteins, the enzymes that passivate them, and complementary enzymes isotropically anchored to a spherical-colloid surface which reactivate the proteins. This passivation-activation cycle gives rise to a symmetry breaking wherein a single biocondensate is stabilized near the colloid surface, which in turn exerts a repulsive force on the colloid. We numerically demonstrate that this mechanism can propel micron-sized colloids at speeds of up to a hundred microns per second. This propulsion mode is strongly resistant to Brownian fluctuations and external forces. Propulsion mechanisms based on biomolecular condensates may thus offer a complementary, motor-free route to biological transport.
\end{abstract}

\maketitle

\section{Introduction}\label{sec:introduction} 

Self-propulsion is a crucial capability of living systems~\cite{MiyataGC2020}. At the cellular scale, motility is typically achieved by means of motor proteins, such as flagellar motors~\cite{WadhwaNRM2022,NakamuraBM2019}, kinesin, dynein, and myosin~\cite{SweeneyCSH2018}, which harness chemical-potential gradients or energy stored in ATP to perform mechanical work~\cite{NojiNature1997}. However, the ability to convert chemical energy into mechanical energy is not exclusive to structurally complex biomolecular machines, finely tuned by evolution. There has been considerable recent interest, across scales, in artificially generating motorless propulsion~\cite{BechingerRMP2016}, as seen in enhanced diffusion of enzymes~\cite{ScottNL2026,GhoshARCMP2021} and self-propelling Janus particles and squirmers~\cite{HowsePRL02007,ThutupalliNJP2011,EbbensSM2010}. Impressively, autonomous propulsion of synthetic isotropic particles is also possible~\cite{MichelinARFM2023}, \eg of diffusiophoretic catalytic droplets~\cite{IzriPRL14,IzzetPRX2020} and actin-nucleating colloids~\cite{BernheimGroswasserNature02,ColinEMBO23}. Directed motion is created by symmetry breaking~\cite{MichelinPF2013,KawskaPNAS12}, thus requiring only a slight perturbation for direction selection instead of, \eg requiring a salt concentration gradient or anisotropic irradiation for a photosensitive system. In this work, we propose a symmetry-breaking self-propulsion mechanism based on simple biological components under physiological conditions.

A natural setting for exploring biologically realizable minimal self-propulsion is active phase separation. Living cells routinely take advantage of this type of physics: Biomolecular condensates serve critical functions in cells~\cite{ShinSCI17,BananiNRCB17}, ranging from signaling~\cite{SuSCI2016} to sequestering cellular components~\cite{MolliexCELL2015} to mediating reactions~\cite{FericCELL2016}. Their location and size are crucial for the normal function of cells~\cite{LeeNP2023,MartinezCalvoPRX25,BuchnerJCP13,CastellanaNB14}, which is why their dynamics are closely regulated by active processes~\cite{ZwickerROPP25,ZwickerPRE15,ZwickerPNAS2014,SorkinJACS2025,HeNCB2026}. Two phenomena in particular suggest that active condensates can also be exploited for transport: (i) Condensates are able to locally rearrange the architecture of the cell by applying physical forces of up to piconewtons~\cite{StromCell2024,SemeigazinARXIV25}. (ii) Active processes are used to localize droplets to desired locations via, \eg autocatalysis~\cite{SuSCI2016,SodingTCB2020} or light-induced activation~\cite{StromCell2024,BrumbaughNC24}. 

In this work, we demonstrate that active biomolecular condensates can generate self-propulsion of a micron-sized object. We only require three components---a phase-separating protein, an enzyme passivating it, and a spherically symmetric catalytic colloid that reactivates the protein. Indeed, the homogeneous, isotropic, and autonomous equations of motion predict a stable symmetry-broken state, wherein a condensate forms in the vicinity of the colloid. As the colloid is pushed away by the repulsive condensate, the catalytic activity of the colloid will autonomously keep the condensate in its vicinity. If the condensate forms near an immobilized colloid, the condensate applies repulsive forces as large as a few nanonewtons. Upon releasing the colloid, the propulsion is limited by protein diffusion, which is nevertheless rapid enough to propel the colloid at velocities exceeding tens of microns per second. The effects of other forces acting upon the colloid, such as Brownian motion or external forces up to a nanonewton, are strongly suppressed. We thus conclude that active biomolecular condensates can propel catalytic structures via a minimal mechanism involving simple biochemical interactions.

The structure of the paper is as follows: In Sec.~\ref{sec:model}, we conceptualize active condensate-mediated propulsion and outline the computational model. In Sec.~\ref{sec:state}, we study the phase diagram emerging from the model and compute the net forces applied to an immobile colloid. In Sec.~\ref{sec:transport}, we compute the drift velocities of a mobile colloid and study the emergent resistance to diffusion and external forces. In Sec.~\ref{sec:p-condensate}, we present a proof-of-concept demonstration of a condensate-propelled condensate. In Sec.~\ref{sec:discussion}, we discuss our results and assumptions, compare the results with conventional self-diffusiophoresis, and connect to general nonequilibrium-thermodynamic principles. Simulation details and parameter values are provided in Appendix~\ref{sec:simulation}, and supporting analytical calculations are given in Appendix~\ref{sec:surface}.

\section{Model}\label{sec:model}

Consider a passive micron-sized colloid, immersed in a passive solvent, \eg water. We seek a minimal set of solutes that will lead to sustained propulsion of the colloid. We require the colloid to be spherically symmetric and corresponding equations of motion to be isotropic, homogeneous, and autonomous, such that a random propulsion direction is selected via spontaneous symmetry breaking. Motivated by recent works---showing that biomolecular condensates (i) are able to apply forces up to a $\mathrm{pN}$~\cite{StromCell2024} and (ii) their positions can be manipulated by (external) activity~\cite{StromCell2024,BrumbaughNC24,SodingTCB2020}---we opt for phase separation as the source of symmetry breaking. Note that we aim for self-organized motion (without the need for external forcing), so the active mechanism should be designed to autonomously localize a phase-separated droplet to the vicinity of the colloid. The proposed minimal components are summarized below and illustrated in Fig.~\ref{fig:illust}(a).

\subsection{Construction}

We first detail the passive components of our model. In addition to a single, spherical colloid, we dissolve ``sticky'' proteins---proteins with a propensity to phase separate. As such, the free energy underlying the proteins' thermodynamics entails two stable phases in equilibrium---a dense (sticky-rich) phase and a dilute (sticky-poor) phase~\cite{DoiBOOK13}. To allow the coexistence of a dense droplet within a dilute background, the solution must be supersaturated\,---\,the total concentration of proteins $\rho_\tot$ must lie above the dilute-phase coexistence concentration. The mechanism we demonstrate in this work does not require high supersaturation: Indeed, we consider systems with $\rho_\tot$ within the binodal region---above the dilute-phase binodal concentration $\rho_\bin$ but below the spinodal concentration $\rho_\spn$.

Without much loss of generality, we assume that the proteins are repelled from the colloid due to, \eg entropic exclusion~\cite{JoannyJPS79,FleerBOOK98} and, in specific cases, electrostatic repulsion~\cite{FleerBOOK98}. Therefore, as a condensate is brought into proximity with the colloid, the two will mutually repel due to molecular repulsion and capillary forces (arising in case the condensate deforms). As a passive solution, the sticky proteins alone are insufficient to generate persistent colloid propulsion since this would constitute an unphysical ``perpetual motion machine.'' Because of the mutual repulsion, the droplet and colloid would localize far from each other at equilibrium. Therefore, our next aim is to introduce chemical activity to autonomously localize the droplet to the sphere.

For the active part of the model, we introduce interconversion reactions (\eg a phosphorylation cycle). We dissolve an enzyme (\eg a kinase) that modifies the sticky species into another state with a weaker phase-separation propensity, and the complementary enzyme (\eg a phosphatase) that restores the original sticky state. For simplicity, we take the other protein state to be ``nonsticky''---a protein that is not prone to phase separation at all (\eg phosphorylation often disrupts phase separation~\cite{HeNCB2026,McIntyreNC25}) and is assumed equivalent to the implicit solvent.

Only dissolving one enzyme is insufficient\,---\,with just a kinase, there will be no sticky proteins to drive phase separation at steady state, while with just a phosphatase, we recover the passive system. Homogeneously dissolving both enzymes will not enable transport either\,---\,instead, size control may occur~\cite{ZwickerPRE15,ZwickerROPP25}, wherein multiple droplets of similar radii coexist at steady state. By our assumption of repulsion, these will localize away from the colloid. 

To maintain one or more droplets in proximity to the colloid, we propose to homogeneously dissolve the kinase but isotropically coat the sphere with the phosphatase. This process would localize sticky-protein production to the vicinity of the colloid and, therefore, if sufficient material is available, allow the formation of droplets close to the colloid surface. In the case of a single droplet, propulsion of the colloid can occur: As the droplet pushes on the colloid, the latter displaces and, to keep up, the bulk kinase decomposes the droplet far from the colloid while the phosphatases allow the growth of the droplet near the sphere, thereby generating self-sustained propulsion. In the remainder of this work, we present evidence supporting this proposed mechanism, illustrated in Figs.~\ref{fig:illust}(b)--\ref{fig:illust}(d).

\begin{figure*}
    \centering
    \includegraphics[width=0.99\linewidth]{illust.pdf}
    \caption{Model for droplet-mediated transport of a micron-sized colloid. (a) Illustration of the model components. Sticky proteins are dissolved in solution above the binodal concentration and thus undergo phase separation. Propulsion is enabled via a phosphorylation cycle: A kinase phosphorylates the sticky proteins, rendering them nonsticky. To localize droplet formation near the colloid, the colloid surface is coated with a phosphatase, which dephosphorylates nearby proteins and restores their stickiness. (b)--(d) Illustration of the pre-steady-state dynamics. (b) Sticky and nonsticky proteins with concentration profiles corresponding to the chemical steady state of the phosphorylation cycle upon dissolving all components. (c) Nucleation occurs in the vicinity of the colloid, where the proteins are predominantly in the sticky state. (d) Single-droplet state (with a spontaneously selected orientation), which is stable under the phosphorylation cycle, in a parameter regime specified later. This asymmetry generates a net force on the colloid, leading to propulsion. (e)--(h) Typical steady-state configurations observed in different parameter regimes. As the overall saturation is increased and/or the phosphorylation rate is decreased, the number of stable droplets increases. (e) No stable droplet. (f) A single stable droplet [as in panel (d)]. (g) Two stable droplets. (h) Six stable droplets. We show three-dimensional renderings of the four steady-state configurations around an immobile colloid in the Supplemental Material~\cite{SM}.}
    \label{fig:illust}
\end{figure*}

\subsection{Equations of motion}\label{sec:PDEs}

With this setup in mind, we formulate dynamical equations describing a mixture of sticky (and nonsticky) proteins, kinase, and a phosphatase-coated colloid. The proteins are nanometer sized and thus much smaller than the micron-sized colloid. Therefore, we consider the continuum limit, viewing the system in terms of sticky- and nonsticky-protein concentration fields, $\rho_\s(\vecr,t)$ and $\rho_\ns(\vecr,t)$~\cite{BrayAP94}. It will prove convenient to work in the colloid's moving reference frame; we denote its velocity by $\vecv(t)$ and position by $\vecq(t)=\int_0^t\d t'\vecv(t')$.

We treat the solvent implicitly. We assume that the local concentration of biomolecules in the solution is subject to strong osmotic constraints, meaning that the total concentration of proteins is capped everywhere by a constant maximal density, $\rho_\max$. We model the bulk properties of the sticky and nonsticky proteins using a Flory-Huggins-type free-energy density~\cite{FloryJCP42,HugginsJCP41},
\begin{multline}
    \frac{g(\rho_\s,\rho_\ns)}{\kt}=\rho_\s\ln\rho_\s+\rho_\ns\ln\rho_\ns\\+(\rho_\max-\rho_\s-\rho_\ns)\ln(\rho_\max-\rho_\s-\rho_\ns)-\frac{\chi}{\rho_\max}\rho_\s^2,\label{eq:FH}
\end{multline}
where the first three terms are the ideal-mixture entropy, and the last term is the attraction energy between sticky proteins. The net attraction strength $\chi$, relative to the implicit solvent and nonsticky proteins, is normalized by the maximum concentration $\rho_\max$ and is thus dimensionless. For large enough $\chi$, Eq.~\eqref{eq:FH} gives rise to a sticky-rich (dense) and a sticky-poor (dilute) phase. The dilute-phase binodal and spinodal concentrations are approximately $\rho_\bin/\rho_\max\sim e^{-\chi}$ and $\rho_\spn/\rho_\max\sim1/(2\chi)$ for $\chi\gg1$. 

To properly model nonuniform density concentration fields around the colloid, we extend the above treatment to a free-energy functional. It contains two contributions: one from the phase separation of sticky proteins, which is captured via a Cahn-Hilliard-type free energy~\cite{CahnJCP58,MaoSM19}, consisting of the bulk free energy of Eq.~\eqref{eq:FH} along with a dense-dilute interfacial term stemming from sticky-sticky interactions; the other from the entropic or electrostatic repulsion of proteins from the colloid~\cite{FleerBOOK98}, which is only a function of distance from its center as the colloid is spherically symmetric. Combined, the free-energy functional is
\begin{multline}
    \frac{G[\rho_\s,\rho_\ns]}{\kt}=\int \d\vecr \biggl\{\frac{g(\rho_\s(\vecr),\rho_\ns(\vecr))}{\kt}+\frac{\chi\xi^2}{\rho_\max}|\grad\rho_\s(\vecr)|^2\\+[\rho_\s(\vecr)+\rho_\ns(\vecr)]\frac{U(|\vecr|)}{\kt}\biggr\},\label{eq:CH}
\end{multline}
where $\xi$ is the dense-dilute interface width and $U(r)$ is the short-ranged repulsion between the colloid surface and a single protein. As we mentioned above, we work in the colloid's moving reference frame, so we arbitrarily placed it at the origin in the protein-colloid repulsion term.

Prior to introducing the reactions, the proteins' passive dynamics is conservative. In the moving reference frame, $\partial\rho_i/\partial t=-\grad\cdot\vecJ_i+\vecv\cdot\grad\rho_i$ (with $i=\s,\ns$). With the free energy of Eq.~\eqref{eq:CH}, the protein flux $\vecJ_i$ drives proteins down chemical-potential gradients~\cite{deGrootBOOK84,BrayAP94}, 
\begin{equation}
    \vecJ_i(\vecr,t)=-\mu_\mol\rho_i(\vecr,t)\grad\frac{\delta G}{\delta\rho_i(\vecr,t)},\qquad i=\s,\ns,\label{eq:flux}
\end{equation}
where $\mu_\mol$ is the molecular mobility constant, assumed to be identical for both protein species. [For completeness, we write the fluxes explicitly in Eqs.~\eqref{eq:flux_expl_s} and~\eqref{eq:flux_expl_ns}.] Similarly, the colloid undergoes Brownian motion subject to the free-energy functional Eq.~\eqref{eq:CH}. In the overdamped limit~\cite{SchussBOOK2010,DoiBOOK13}, its drift is therefore
\begin{equation}
    \vecv(t)=\mu_\col [\vecF(t)+\vecF_\ext]+\sqrt{2\kt\mu_\col}\boldsymbol\zeta(t),\label{eq:drift}
\end{equation}
where $\mu_\col$ is the colloidal mobility constant, the force $\vecF=-\partial G/\partial\vecq|_{\vecq=\mathbf0}$ is
\begin{equation}
    \vecF(t)=\int\d\vecr[\rho_\s(\vecr,t)+\rho_\ns(\vecr,t)]\grad U(|\vecr|),\label{eq:force}
\end{equation}
$\vecF_\ext$ is any deterministic external force, and the Brownian white noise satisfies $\langle\boldsymbol\zeta(t)\boldsymbol\zeta(t')\rangle=\boldsymbol1\delta(t-t')$. Both mobilities are known from the Stokes-Einstein relation~\cite{EinsteinAP1905,DoiBOOK13}.

We now introduce the activity. Assuming the fuel (\eg ATP) concentration is fixed, we take the degradation reactions to follow first-order kinetics, $\partial\rho_\s/\partial t=-\partial\rho_\ns/\partial t=-k\rho_\s$, with a constant phosphorylation rate $k$. As we argue in Appendix~\ref{sec:surface}, for sufficiently fast dephosphorylation, we can assume that all nonsticky proteins that reach the surface of the colloid are instantaneously converted to sticky, so that
\begin{equation}
    \rho_\ns(|\vecr|=R,t)=0,\label{eq:BCinstantconv}
\end{equation}
where $R$ is the radius of the colloid. In addition, we must impose no-flux boundary conditions perpendicular to the colloid's surface, 
\begin{equation}
    J_{\s,r}(|\vecr|=R,t)+J_{\ns,r}(|\vecr|=R,t)=v_{r}(t)\rho_\s(|\vecr|=R,t).\label{eq:BCnoflux}
\end{equation}

Combining the above, the full equations of motion are
\begin{align}
    \frac{\partial\rho_\s(\vecr,t)}{\partial t}=&-\grad\cdot\vecJ_\s(\vecr,t)+\vecv(t)\cdot\grad\rho_\s(\vecr,t)\nonumber\\&-k\rho_\s(\vecr,t),\label{eq:PDE_s}\\
    \frac{\partial\rho_\ns(\vecr,t)}{\partial t}=&-\grad\cdot\vecJ_\ns(\vecr,t)+\vecv(t)\cdot\grad\rho_\ns(\vecr,t)\nonumber\\&+k\rho_\s(\vecr,t),\label{eq:PDE_ns}
\end{align}
with the described boundary conditions [Eqs.~\eqref{eq:BCinstantconv} and~\eqref{eq:BCnoflux}], fluxes [Eq.~\eqref{eq:flux}, or Eqs.~\eqref{eq:flux_expl_s} and~\eqref{eq:flux_expl_ns}], and colloidal drift [Eq.~\eqref{eq:drift}]. Indeed, these equations are homogeneous (which we have taken advantage of by working in the colloid's moving frame), isotropic, and autonomous. The type of chemical activity we introduced is also minimal.

In the remainder of this work, we show and discuss the results of numerical simulations of Eqs.~\eqref{eq:PDE_s} and~\eqref{eq:PDE_ns}. Simulation details, discussion of the equations' range of validity, and parameter values are detailed in Appendix~\ref{sec:simulation}. We systematically scan through $\rho_\tot$, the total concentration of proteins (which is distributed between the sticky and nonsticky forms in response to the reactions), and $\ell_k=(\mu_\mol\kt/k)^{1/2}$, the average distance diffused by a sticky protein until phosphorylation, and observe the fate of a droplet initially seeded on one side of the condensate. Varying these two key parameters sheds light on the physics behind the emergent behavior, including autonomous propulsion. For brevity, we only show the results for two colloid radii, $R=0.2\micron,0.5\micron$, and keep other parameters fixed.

\section{Immobile colloid}\label{sec:state}

\begin{figure}
    \centering
    \includegraphics[width=0.99\linewidth]{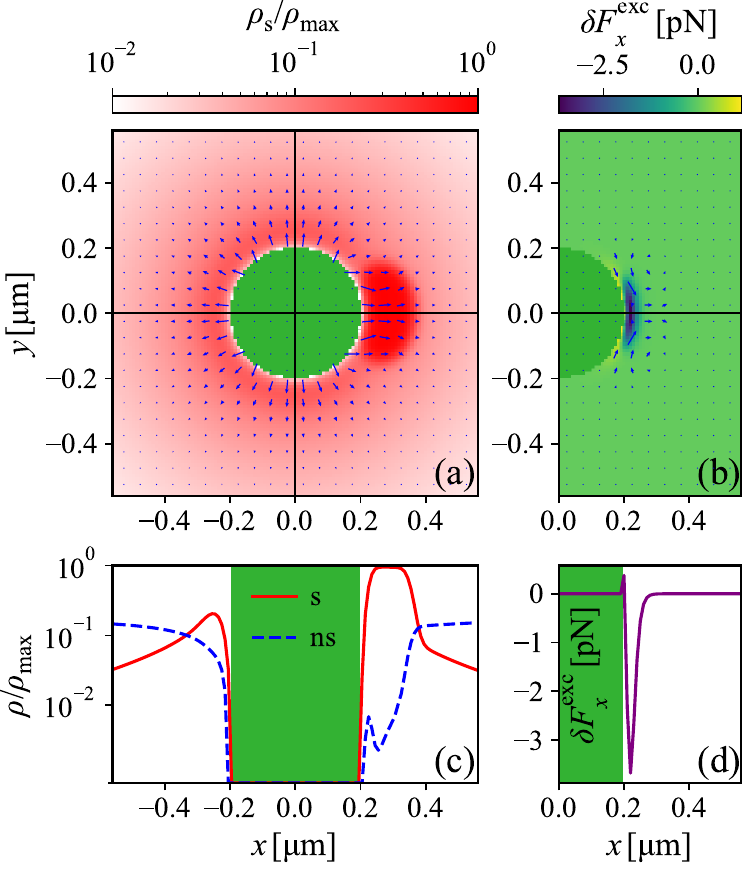}
    \caption{Three-dimensional simulations of an immobilized colloid. (a),(b) Two-dimensional cross section through the center of the colloid (green) and the droplet. (a) Sticky-protein concentration profile, $\rho_\s(x,y,z=0)$ (red, normalized by the maximal density, $\rho_\mathrm{max}=0.1\mathrm{nm}^{-3}$). The ``pixels'' exactly represent our finite-volume density elements. Blue arrows show the sticky-protein flux $\vecJ_\s(x,y,z=0)$ as a guide, computed on a $5\times5$-fold coarser grid (for clarity), where the arrow lengths are scaled such that the largest $|\vecJ|$ is five pixels ($0.05\mathrm{nm}$). (b) Excess force along the $x$ direction, defined as the difference between the forces exerted by proteins on the right and left sides of the colloid, $\delta F_x^\mathrm{exc}(x,y,z=0)$. Blue arrows show the excess sticky-protein flux, defined in a similar manner, $(J_x(x,y,0)+J_x(-x,y,0),J_y(x,y,0)-J_y(-x,y,0))$, and normalized again so the longest arrow is five pixels long. (c),(d) Line scan along the line connecting the centers of the colloid and the droplet of (c) the sticky and nonsticky concentration profiles, $\rho_i(x,y=0,z=0)$ ($i=\s,\ns$), and (d) the excess force, $\delta F_x^\mathrm{exc}(x,y=0,z=0)$. Parameters: colloid radius $R=0.2\micron$; average distance between phosphorylations $\ell_k=0.64\micron$; and supersaturation relative to the spinodal-binodal difference $(\rho_\tot-\rho_\bin)/(\rho_\spn-\rho_\bin)=0.75$. These plots are close-ups; the total simulated-system size is $(4.8\micron)^3$, with periodic boundary conditions.}
    \label{fig:force_profile}
\end{figure}

\begin{figure*}
    \centering
    \includegraphics[width=0.99\linewidth]{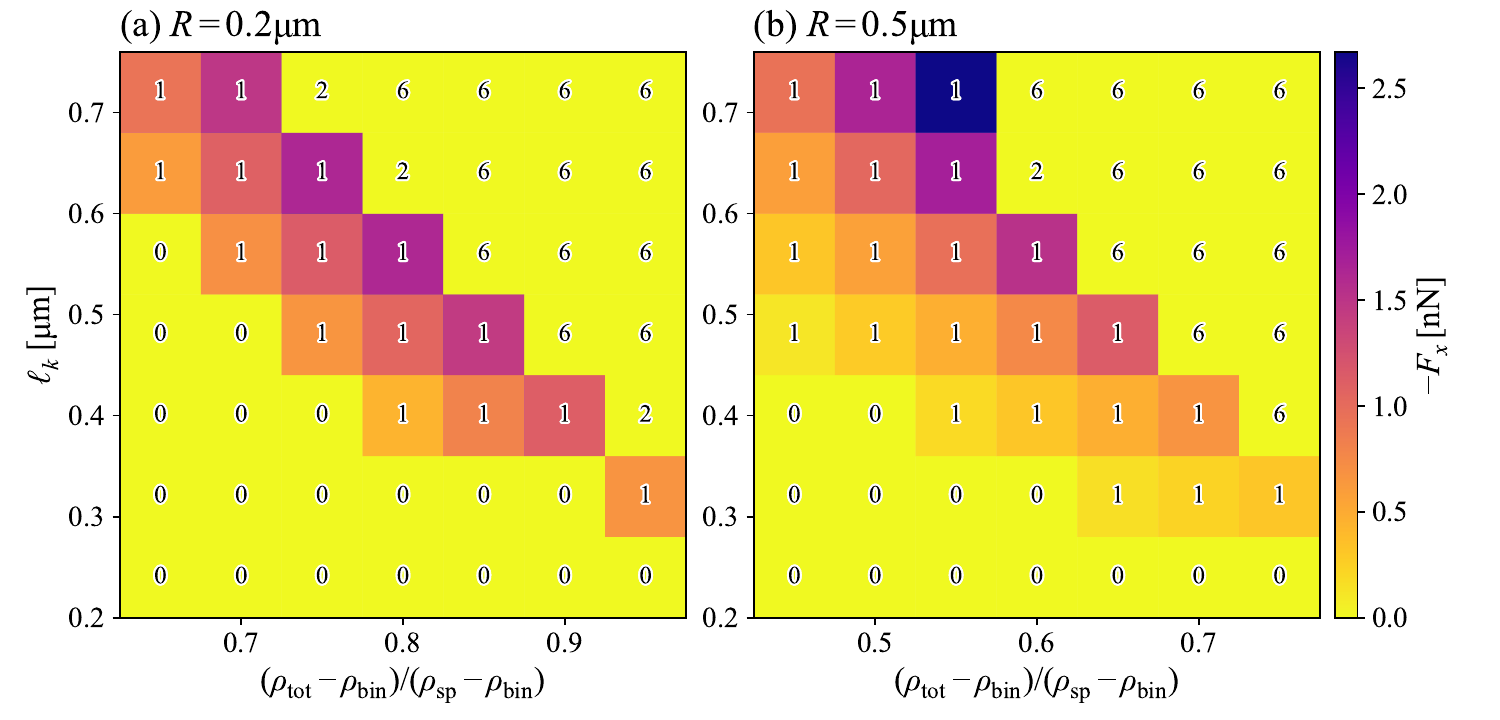}
    \caption{Phase diagrams for three-dimensional simulations of an immobilized colloid with radii (a) $R=0.2\micron$ and (b) $R=0.5\micron$. The steady-state number of stable droplets [\cf Figs.~\ref{fig:illust}(e)--(h)] is indicated at the center of each square, and the magnitude of the leftward force in single-droplet configurations is represented by color. The control parameters are the average distance until phosphorylation, $\ell_k$, and the supersaturation relative to the spinodal-binodal difference, $(\rho_\tot-\rho_\bin)/(\rho_\spn-\rho_\bin)$. Increasing the ambient concentration and/or decreasing the phosphorylation rate permits a larger number of stable droplets, and furthermore increases the net force in the single-droplet regime. We show three-dimensional renderings of the four steady-state configurations in the Supplemental Material~\cite{SM}.}
    \label{fig:diagram_static}
\end{figure*} 

\subsection{Steady configurations}

We begin with simulations of an immobilized colloid [$\mu_\col=0$ in Eq.~\eqref{eq:drift}, so $\vecv=\mathbf0$ in Eqs.~\eqref{eq:PDE_s} and~\eqref{eq:PDE_ns}]. We scan through the total protein concentration $\rho_\tot$ and the inter-phosphorylation distance $\ell_k$, and observe four types of steady states. Although $\rho_\tot\geq\rho_\bin$ in all our simulations, most of the proteins are in a nonsticky state due to the action of the kinase in the bulk, causing the sticky profile to fall off exponentially away from the colloid over a distance $\ell_k$ (see Appendix~\ref{sec:surface-id}). Thus, for lower $\rho_\tot$ and/or shorter $\ell_k$, there is insufficient sticky material to maintain a droplet. Indeed, an initially placed droplet dissolves, and the system reaches a no-droplet steady state (Fig.~\ref{fig:illust}(e)). 

However, if we add more proteins (bigger $\rho_\tot$) or remove some kinase (bigger $\ell_k$), a steady state with a single stable droplet is reached [Fig.~\ref{fig:illust}(f) and Fig.~\ref{fig:force_profile}(a)]. Notably, this single droplet is not spherical but rather ``cap'' shaped. If we keep increasing $\rho_\tot$ and $\ell_k$, there is sufficient sticky material around the colloid to form additional droplets, so we observe two [Fig.~\ref{fig:illust}(g)] and even six [Fig.~\ref{fig:illust}(h)] droplets. The number of droplets at steady state is indicated at the center of each cell in Fig.~\ref{fig:diagram_static} for two colloid sizes. It is important to note that the system exhibits multistability---initial conditions of different symmetries can lead to different steady states. Since Eqs.~\eqref{eq:PDE_s} and~\eqref{eq:PDE_ns} are noiseless, in order to break spherical symmetry, the results of Fig.~\ref{fig:diagram_static} originated from simulations in which we initially seeded a single droplet to the right of the colloid (see Appendix~\ref{sec:simulation-coll}).

The emergence of droplets can be understood through the joint action of the phosphatase-coated colloid and the bulk activity of the kinase. We first unpack the single-droplet morphology. Once a threshold curve in $\rho_\tot$--$\ell_k$ parameter space is crossed, the phosphatase creates a wide enough ``shell'' of sticky material to permit a stable droplet; see Fig.~\ref{fig:force_profile}(a). Note that since the droplet is situated between a (repulsive) colloid, into which it may not grow, and a kinase-ridden bulk that decomposes sticky material, it may not coarsen beyond a width of around $\ell_k$. Instead, we reach a steady situation where sticky material is being added to the droplet from the colloid's surface and being decomposed on its other side, thereby giving the droplet a caplike shape. This shape is thermodynamically unfavorable, as droplets would prefer to be spherical to minimize surface tension; nevertheless, activity maintains this state indefinitely.

Upon increasing $\rho_\tot$ and $\ell_k$ further, why do multidroplet states appear? In principle, a single droplet would be thermodynamically favored to minimize surface area~\cite{LifshitzJPCS61,SiggiaPRA79,CatesBOOK17,EralLangmuir11,SmithCS81}. Despite this fact, upon crossing another $\rho_\tot$--$\ell_k$ threshold curve, we find a steady state consisting of two droplets, rather than a further increase in the size of the single droplet. A key factor stabilizing the two-droplet state is that as we increase, \eg $\ell_k$, the sticky-shell width grows as $\ell_k$, and the integrated amount of sticky material in the shell grows as $\ell^2_k$, so there is enough sticky material to support a second droplet. From symmetry considerations, at steady state, the second droplet always sits opposite the first droplet, and reaches exactly the same size. The two-droplet regime is rather narrow: As we further increase $\rho_\tot$ and $\ell_k$, we find a steady state with more (six) identically sized droplets.\footnote{Interestingly, at no point did we observe three to five droplets at a steady state because (i) these states are truly always transient states, (ii) the corresponding range of parameters is extremely narrow, or (iii) this is a shortcoming of finite-volume simulations on a cubic lattice.} Overall, these configuration are a direct consequence of activity and would not occur in equilibrium.

\subsection{Forces and concentration fields}

Having described the steady states attainable for this system, we proceed to characterize the forces the colloid experiences. In Fig.~\ref{fig:diagram_static}, we color code the force magnitudes on the sphere due to the proteins [Eq.~\eqref{eq:force}]. Unsurprisingly, because of reflection symmetries in all three dimensions, the no-, two-, and six-droplet states do not exert a net force. However, the single-droplet state applies a net force on the colloid in the direction opposite to the droplet position ($\vecF=F_x\unit{x}$, $F_x<0$). Because of the reflection symmetry, the net force along the $\unit{y}$ and $\unit{z}$ directions is zero. This finding suggests that, that upon releasing the colloid, it may be pushed into motion by the droplet; we confirm that this is the case in Sec.~\ref{sec:transport}. 

Prior to characterizing the force magnitudes, we first inspect the concentration profiles for the single-droplet configuration for $R=0.2\micron$.\footnote{The ``pixels'' in Figs.~\ref{fig:force_profile}(a) and~\ref{fig:force_profile}(b) exactly represent our finite-volume density elements and, on a staggered lattice, the $\unit{x}$-ward force elements, respectively.} In Fig.~\ref{fig:force_profile}(a), we plot the $z=0$ cross section of the sticky-protein concentration. A red-colored dilute-phase ``halo'' of sticky proteins is seen around the colloid, which is sufficient to stabilize a single droplet to the right of the colloid. Blue arrows represent the $\unit x$ and $\unit y$ components of the sticky-protein flux, $\vecJ_\s(x,y,0)$, showing how the surface-made sticky material diffuses outwards, and how the droplet warps the sticky-protein flow lines. [The fluxes are large in magnitude within the dense phase as a result of concentration-gradient contributions to the flux being amplified by the entropic penalty $\rho_\max-\rho_\s-\rho_\ns$ in the denominator of Eqs.~\eqref{eq:flux_expl_s} and~\eqref{eq:flux_expl_ns}.]
In Fig.~\ref{fig:force_profile}(b), we plot the excess leftward force exerted on the colloid, $\delta F_x^\mathrm{exc}(x,y,z=0)=\d x\d y\d z[\rho_\s(x,y,0) (\partial/\partial x)U(x,y,0)-\rho_\s(-x,y,0)(\partial/\partial x) U(-x,y,0)]$, which is simply the integrand times the volume element in Eq.~\eqref{eq:force} at position $(x,y,0)$, minus the corresponding value at $(-x,y,0)$. The positive force elements (net rightward push on the colloid) immediately surrounding the condensate result from the lower local concentration of around $\rho_\bin$---the dilute-phase coexistence concentration---at the droplet interface, compared to the ambient concentration of around $\rho_\tot$ on the other side of the colloid. However, the bulk of the condensate applies a strong enough leftward force to overall push the sphere strongly to the left. The blue arrows in this panel---representing the excess flux elements to the right of the colloid versus to its left, $(J_x(x,y,0)+J_x(-x,y,0),J_y(x,y,0)-J_y(-x,y,0))$---show how the droplet is maintained at a steady state: Sticky material escapes the droplet from its right, but joins through an influx tangential to the colloid's surface.

In Fig.~\ref{fig:force_profile}(c), we plot a close-up on the $z=y=0$ line (passing through the centers of the colloid and droplet) of the sticky- and nonsticky-protein concentrations. They vividly show the sticky condensate to the colloid's right, the exclusion of the nonsticky species from the condensate, the rapid decrease in both concentrations close to the colloid surface, and the exponential decay of the sticky concentration (and the complementary rise in the nonsticky concentration) away from the colloid. It is those profiles that keep the droplet near the colloid surface. For completeness, we show a similar plot for the excess force in Fig.~\ref{fig:force_profile}(d). It is the protein concentration imbalance between the right and left of the colloid that leads to an overall leftward force.

Now, with a mechanistic understanding of the concentration profiles and how they generate a force, we return to Fig.~\ref{fig:diagram_static} and inspect the force magnitudes inside the single-droplet, ``propulsive'' band. Following our previous discussion, as we increase $\rho_\tot$ and $\ell_k$ within this band, the single droplet increases in size as more material is available for it. As it does so, the imbalance between protein concentrations on either side of the colloid increases, too, thereby leading to a larger force (which is depicted by the darker colors). Notably, the boundary of the propulsive band is sharp because a change in the number of droplets is a discrete and abrupt process. For example, upon passing from the no- to one-droplet regions, the appearance of a droplet is concomitant with an immediate concentration imbalance, generating a nonzero net force. Similar considerations underlie the crossover from the single- to two-droplet regions: Once the second droplet is present, it immediately negates the force due to the first droplet.

For $\vecv=\mathbf0$, the above phenomena are independent of our choice of $\rho_\max$ or the interface width $\xi$. However, to assign a dimensional value to the force, we must pick a number of proteins in an interface-width voxel; we choose $\rho_\max\xi^3=10^2$ and $\xi=10\mathrm{nm}$ (see Appendix~\ref{sec:simulation-para}). With this order-of-magnitude choice, we note a rather perplexing fact\,---\,the forces in Fig.~\ref{fig:diagram_static} reach a few $\mathrm{nN}$s. This result follows because each molecule in direct vicinity to the colloid repels it with a force reaching up to $\kt/\xi\sim1\mathrm{pN}$ (see Appendix~\ref{sec:simulation-para} for parameter values), and those forces accumulate to around $1\mathrm{nN}$ for a mesoscopic condensate. These forces are $10^4$-fold larger than the capillary forces estimated for synthetic nuclear condensates~\cite{StromCell2024} and are even larger than the forces generated by typical optical tweezers~\cite{NeumanREV04}. According to the Stokes mobility~\cite{DoiBOOK13}, $\mu_\col=1/(6\pi\eta R)$, a force of around $1\mathrm{nN}$ would propel a mobile micron-sized colloid in water with a speed of around $0.1\mathrm{m}/\mathrm{s}$. This force is not physical---because we must treat a mobile colloid self-consistently. 

\begin{figure*}
    \centering
    \includegraphics[width=0.99\linewidth]{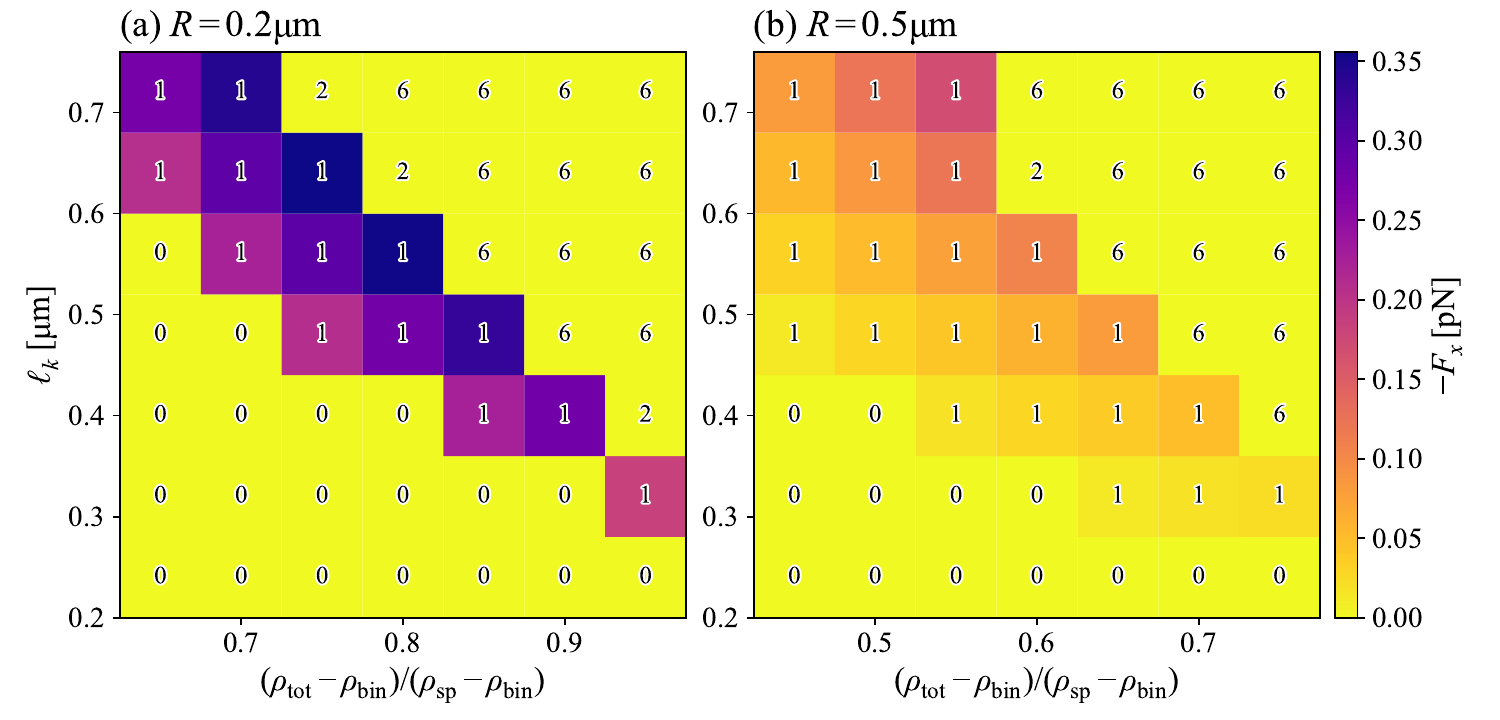}
    \caption{Phase diagrams for three-dimensional simulations of a mobile colloid in water at room temperature. Consequently, the force is dramatically reduced compared to Fig.~\ref{fig:diagram_static}.}
    \label{fig:diagram_free}
\end{figure*}

\section{Mobile colloid}\label{sec:transport}

Since the colloid's propulsion requires that the droplet remains in its vicinity, the colloid's speed is constrained. For nonzero colloid speeds, $|\vecv|>0$, the proteins must be able to reorganize in time for the droplet to ``keep up'': As the colloid displaces leftwards, nonsticky proteins must be able to arrive at its surface via a diffusive flux, convert to a sticky state, flow back towards the droplet and attach to its left side. On the droplet's other side, the kinase must simultaneously passivate the sticky material to decompose the right side of the droplet. We also note that the dilute phase to the left of the colloid must ``clear the way,'' thereby imposing an additional resistance to the drift. As we will show below, at steady state, this balance leads to a joint colloid-droplet drift $\vecv$, which is orders of magnitude smaller than the above estimate (obtained from the immobilized-colloid forces). This balanced state is self-correcting: For example, if $\vecv$ is initially too large, the droplet would lag behind, and $\vecv$ would keep slowing down until the stable droplet configuration is reached. Below, we consider three cases of a mobile colloid: a nondiffusing free colloid, a Brownian free colloid, and a nondiffusing driven colloid.

\subsection{Non-Brownian colloid}\label{sec:nonBro}

We repeat the simulations of Eqs.~\eqref{eq:PDE_s} and~\eqref{eq:PDE_ns} for a free colloid, whose mobility in Eq.~\eqref{eq:drift} is given by the Stokes formula~\cite{LandauFluid1987}, $\mu_\col=1/(6\pi\eta R)$, where $\eta$ ($=10^{-3}\mathrm{Pa}\cdot\s$) is the solvent (water) viscosity. For this section, we set Brownian noise to zero [$\zeta(t)=0$]. We scan through total concentration $\rho_\tot$ and inter-phosphorylation distance $\ell_k$. We again observe the same four types of steady states arranged across the same regions.\footnote{The phase diagram has only changed at $((\rho_\tot-\rho_\bin)/(\rho_\spn-\rho_\bin)=0.9,\ell_k=0.32\micron)$, where the single droplet has destabilized.} Indeed, the no-, two-, and six-droplet cases apply no net force on the colloid, so once the system reaches a steady state (after starting with a spherical droplet on the right), the velocity is $\vecv = \mathbf0$. For a single droplet, however, as the protein concentrations rearrange in response to the colloid's motion, the force persists but is reduced dramatically, compared to Fig.~\ref{fig:diagram_static}, to a fraction of a $\mathrm{pN}$. Using the Stokes mobility, this process nonetheless amounts to a considerable propulsion velocity, $\mathcal{O}(10\micron/\s)$.

\subsection{Brownian colloid}\label{sec:Bro}

For the case of a single droplet propelling a mobile colloid, we next ask how persistent this motion is. Namely, upon reintroducing Brownian motion in Eq.~\eqref{eq:drift}, we assess the resistance of the directed motion to random reorientation due to thermal noise. To estimate the importance of diffusion relative to propulsion, we consider the P\'eclet number at the scale of the colloid,
\begin{equation}
    \mathrm{Pe}\equiv\frac{|\vecv|R}{\mu_\col\kt}=\frac{|\vecF|R}{\kt}.\label{eq:Pe}
\end{equation}
For a mobile colloid experiencing about $0.1\mathrm{pN}$, promisingly, the P\'eclet number is rather large, $\mathrm{Pe}\gtrsim10$. Hence, diffusion is not dominating at any scale. This finding implies that the propulsion's ``persistence length'' can be large.

\begin{figure}
    \centering
    \includegraphics[width=0.99\linewidth]{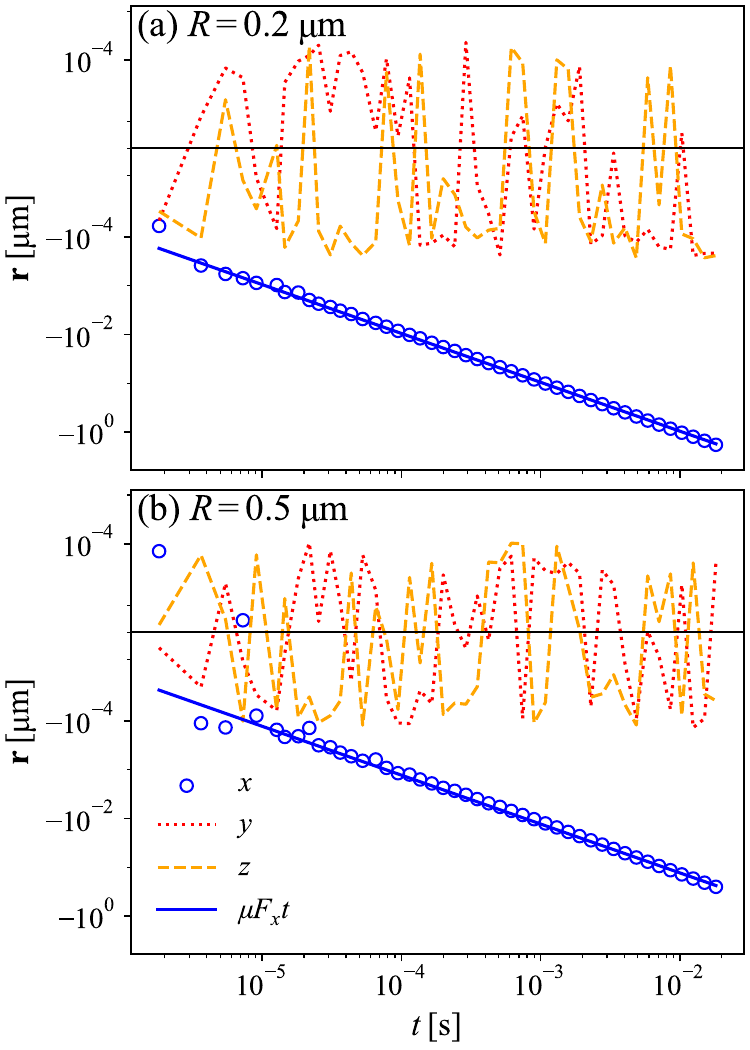}
    \put(-211,179){\includegraphics[width=0.33\linewidth]{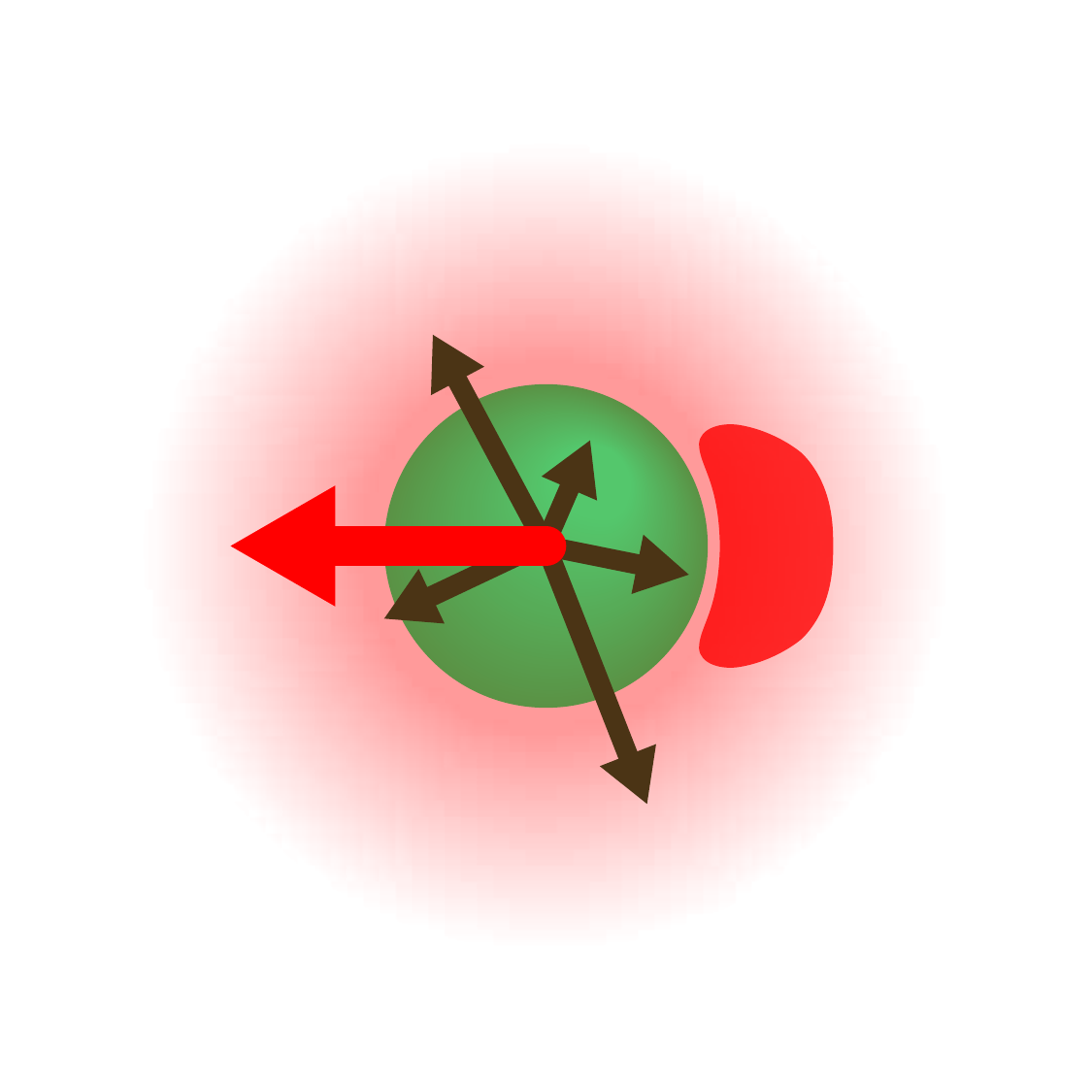}}
    \caption{Three-dimensional simulations of a ``Brownian'' colloid in water at room temperature. We plot a representative stochastic trajectory $\vecr=(x,y,z)$ of a colloid on a symmetric logarithmic scale, for single-droplet conditions, (a) $R=0.2\micron$, $\ell_k=0.64\micron$, and $(\rho_\tot-\rho_\bin)/(\rho_\spn-\rho_\bin)=0.75$, and (b) $R=0.5\micron$, $\ell_k=0.64\micron$, and $(\rho_\tot-\rho_\bin)/(\rho_\spn-\rho_\bin)=0.55$. Inset: illustration of the single-droplet configuration, subject to Brownian noise (brown arrows). The colloid is propelled with a nearly constant drift velocity $\mu F_x$ consistent with the deterministic simulations in Fig.~\ref{fig:diagram_free}.}
    \label{fig:diffusion}
\end{figure}

To go further and include the motion of the colloid relative to the droplet, we simulate Eqs.~\eqref{eq:PDE_s} and~\eqref{eq:PDE_ns} with the full right-hand side of Eq.~\eqref{eq:drift} for a Brownian colloid in water at room temperature ($\mu_\col=1/(6\pi\eta R)$, $\eta=10^{-3}\mathrm{Pa}\cdot\s$, and $\kt=4.1\mathrm{pN}\cdot\mathrm{nm}$), where the initial condition is the non-Brownian steady state (Sec.~\ref{sec:nonBro}). We choose parameters for which a single-droplet configuration emerges prior to adding noise, and we plot the results in Fig.~\ref{fig:diffusion}. Remarkably, we observe indefinite resistance to thermal perturbations. The colloid's position perpendicular to the direction of motion never fluctuates above $0.1\mathrm{nm}$ (dotted red and dashed orange curves). Meanwhile, the stochastic position in the direction of motion ($\unit{x}$, blue circles) self-averages over the course of our simulations of around $10^{-2}\s$ long to exactly the velocities predicted by the non-Brownian simulations of Fig.~\ref{fig:diagram_free}, $|\langle\vecv\rangle|=95\micron/\s$ and $|\langle\vecv\rangle|=13\micron/\s$ for $R=0.2\micron$ and $R=0.5\micron$ colloids, respectively (blue solid lines).

This extreme stability can be understood by recalling the immense forces the immobile colloid experienced (\cf Fig.~\ref{fig:diagram_static}). Since the protein concentration fields adjust to tune the $\mathrm{nN}$s of force down to $\mathrm{pN}$, any fluctuation in the density field may result in restoring forces of the previous $\mathrm{nN}$s. To demonstrate, we view the $y=z=0$ colloid position as the bottom of an effective potential well. Using $\langle y^2\rangle=\langle z^2\rangle\sim(10^{-4}\micron)^2$ and the equipartition theorem, we may estimate an effective spring constant of the restoring potential, $k=\kt/\langle y^2\rangle\sim\mathrm{nN}/(10\mathrm{nm})$. This estimate is indeed on the order of the immobile-colloid forces per unit interface width. 

So far, we have discussed diffusion of the colloid relative to the condensate. However, our simulations do not account for droplet fluctuations. The most direct means to include noise in phase separation would have been to simulate fluctuating field-field equations (see Appendix~\ref{sec:nucl} for discussion). However, since Fig.~\ref{fig:diffusion} implies that any fluctuation of the sphere and droplet away from their steady configuration is strongly suppressed, by omitting condensate noise we are only missing rigid-body-like translational and rotational diffusion of the droplet-colloid pair. (In other words, Fig.~\ref{fig:diffusion} amounts to working in the cotranslational-corotational frame of the condensate-droplet pair.) By noting that $\mathrm{Pe}\gg1$, we already concluded that translational diffusion is negligible compared to propulsion. We thus proceed to inspect rotational diffusion and the resulting persistence length.

We readily find a lower bound on the persistence length as follows: The rotational diffusion of the ``dumbbell'' that is the colloid-condensate pair is bounded from above by the rotational diffusion constant of the spherical colloid~\cite{LandauFluid1987}, $D_\mathrm{rot}=\kt/(8\pi\eta R^3)\sim\mu_\col\kt/R^2$. Thus, $D_\mathrm{rot}^{-1}$ is a lower bound on the typical time for the dumbbell to decorrelate in orientation. The colloid-droplet persistence length is given by the distance traversed during the orientational-decorrelation time, $\ell_\mathrm{p}>|\vecv|/D_\mathrm{rot}\sim \mathrm{Pe}\times R$, where we identified the P\'eclet number of Eq.~\eqref{eq:Pe}. Since $\mathrm{Pe}\gtrsim10$, the persistence length is reassuringly large, being at least on the order of tens of colloidal radii. In conclusion, our simulations demonstrate that the colloid-condensate pair is indefinitely rigidlike. In addition, our theoretical analysis shows that it moves tens of body lengths before reorienting, confirming that the pair can self-propel persistently.

\subsection{Driven colloid}\label{sec:driv}

\begin{figure}
    \centering
    \includegraphics[width=0.99\linewidth]{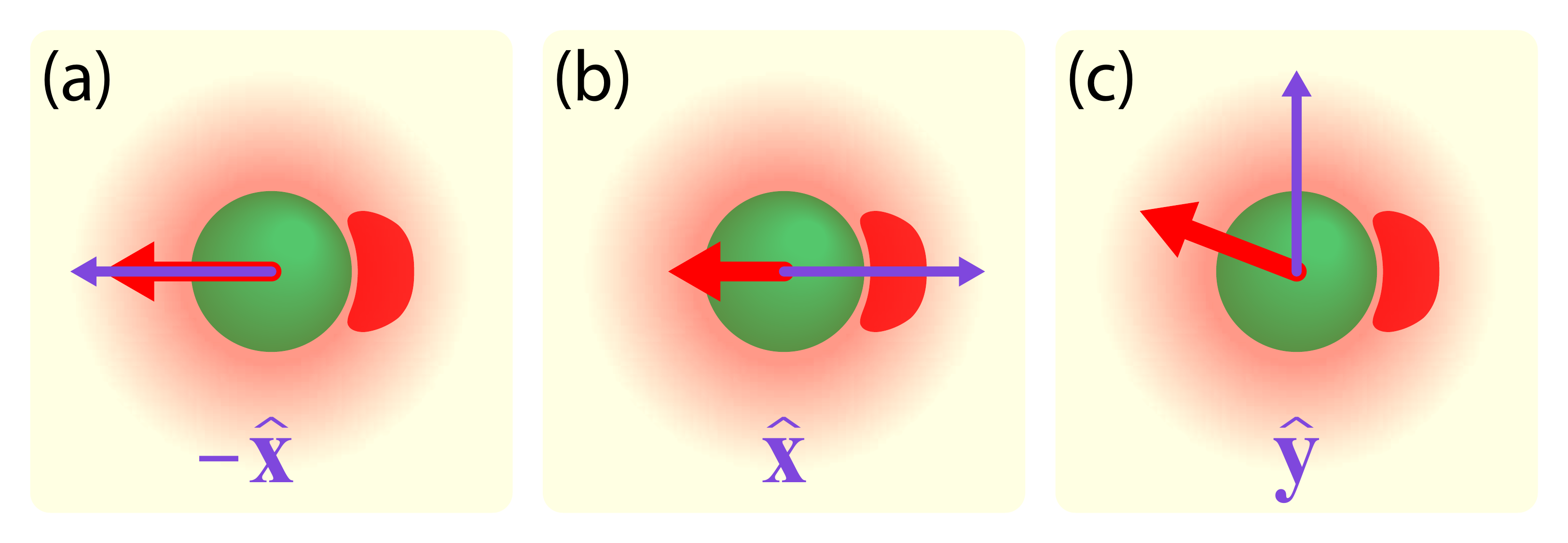}
    \includegraphics[width=0.99\linewidth]{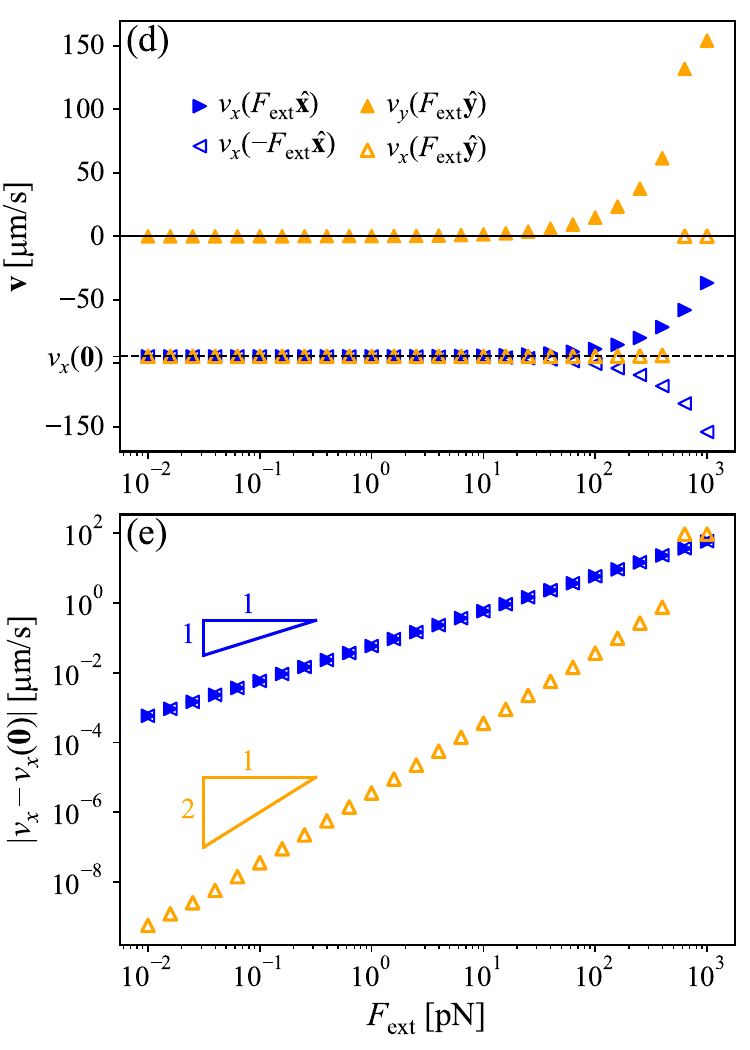}
    \caption{Three-dimensional simulations of an externally driven colloid in water at room temperature. (a)--(c) Illustration of the three external-force directions we consider (purple arrows) and the total drift velocities (red arrows), where the droplet is centered at $x>0$. (d) Colloid drift velocity along the indicated axes as a function of the magnitude of the additional, externally applied force, acting along the specified axes. (e) Magnitude of excess drift velocity relative to the undriven case (Fig.~\ref{fig:diagram_free}) for parallel [blue; panels~(a) and~(b)] and perpendicular [orange; panel~(c)] external forces. Parameters: colloid radius $R=0.2\micron$; average distance before phosphorylation $\ell_k=0.64\micron$; and supersaturation relative to the spinodal-binodal difference $(\rho_\tot-\rho_\bin)/(\rho_\spn-\rho_\bin)=0.75$.}
    \label{fig:drift}
\end{figure}

Having found considerable persistence and stability under Brownian motion, we ignore Brownian motion in this section ($\zeta(t)=0$), and systematically consider the system's susceptibility to deterministic, strong external perturbations. Namely, we drive the colloid with an additional external force. Specifically, we simulate Eqs.~\eqref{eq:drift}, \eqref{eq:PDE_s}, and~\eqref{eq:PDE_ns} with $\boldsymbol\zeta=\boldsymbol0$ and $|\vecF_\ext|>0$ for single-droplet-configuration parameters. We scan external forces within the range $|\vecF_\ext|\in[10^{-2}\mathrm{pN},1\mathrm{nN}]$, corresponding to the range of forces that could be applied by optical tweezers~\cite{NeumanREV04}. We probe the three distinct directions in which we may apply an external force---along the direction of propulsion (leftwards, $-\unit{x}$), opposite it (rightwards, $\unit{x}$), and perpendicular to it (arbitrarily upwards, $\unit{y}$). In Fig.~\ref{fig:drift}(d), we plot the relevant components of the steady drift velocity in response to the force, $\vecv=v_x\unit{x}$, $\vecv=v_x\unit{x}$, and $\vecv=v_x\unit{x}+v_y\unit{y}$, respectively. 

We see that for all three directions of $\vecF_\ext$, the drift velocity remains close to the unperturbed non-Brownian steady drift value for small $|\vecF_\ext|$ [$\vecv=(-95\micron/\s)\unit{x}$ for these parameters; \cf Fig.~\ref{fig:diagram_free}]. At this point, it is not surprising that the region of ``negligible $\vecF_\ext$'' extends rather far, up to $\mathcal{O}(0.1\mathrm{nN})$---the colloid's motion is limited by the ability of the proteins on either side\,---\,in the condensate and in the dilute phase opposite the condensate---to diffusively adjust to its motion. Only when the external forces reach about $0.1\mathrm{nN}$ do their effects become observable---the leftward force increases the leftward drift (left-facing empty blue triangles), the rightward force opposes the leftward drift (right-facing solid blue triangles), and the perpendicular force generates a drift upwards (upward-facing solid orange triangles). Interestingly, in the last scenario, because of the resistance of the proteins to perpendicular perturbations we observed in Sec.~\ref{sec:Bro}, the droplet remains to the right of the colloid and thus keeps pushing on it with approximately the unperturbed leftward force (upward-facing empty orange triangles). Therefore, we may expect that forces such as gravity, which apply about $1\mathrm{fN}$ on micron-sized objects with density exceeding that of water by about $10\%$, will not affect the propulsion. It is only at $\vecF_\ext=(0.79\mathrm{nN})\unit{y}$ that the droplet finally succumbs to the external drive and, before reaching a steady state, tilts to below the colloid.\footnote{Indeed, for not much larger forces in any of the directions, the forces are so large that the droplet destabilizes and dissolves prior to reaching a steady state. The estimate $|\vecv|\sim0.1\mathrm{m}/\s$ is thus (strongly) opposed only by the dilute-phase imbalance on either side of a quickly moving colloid.} In this case, $v_x(F_\ext\unit{y})=0$ and, consistently, $v_y(F_\ext\unit{y})=-v_x(-F_\ext\unit{x})$ (compare the two rightmost left-facing empty blue triangles with the rightmost upward-facing solid orange triangles).

As another indicator of the formidable opposition of the droplet to external drive, in Fig.~\ref{fig:drift}(e) we plot the absolute value of the deviation in leftward velocity for the three external drifts, $|v_x(\vecF_\ext)-v_x(\mathbf0)|$. Consistent with a nonequilibrium linear response~\cite{BaiesiNJP2013}, the leftward drift velocity reacts linearly to external forces parallel to the motion across the entire range of forces considered (blue triangles). Notably, again supporting a linear-response picture, the absolute values of the velocity deviations due to leftward and rightward external forces are equal. Now, in the perpendicular force case, the reduction of the leftward drift is a second-order effect so, unsurprisingly, the leftward drift velocities (orange triangles) decrease quadratically with external force (with a significantly smaller magnitude overall), up until the droplet reorients at $\vecF_\ext=(0.79\mathrm{nN})\unit{y}$. 

\section{Self-propelled condensates}\label{sec:p-condensate}

It may be of interest to confirm that catalytic condensates may also be propelled by the same mechanism---a sticky droplet pushing a catalytic condensate, rather than a colloid. Given recent advances in biocondensate engineering~\cite{DaiNRB23}, it may be simpler to create a phosphatase condensate than to densely coat a colloid with phosphatases. We perform numerical simulations akin to Eqs.~\eqref{eq:PDE_s} and~\eqref{eq:PDE_ns}, where instead of a colloid, we introduce a third concentration field, $\rho_\p(\vecr,t)$, of a phosphatase with a propensity to phase separate.\footnote{The colloid-surface boundary conditions, Eqs.~\eqref{eq:BCinstantconv} and~\eqref{eq:BCnoflux}, are no longer relevant, nor is the drift Eq.~\eqref{eq:drift}. If a propelling state is achieved, we would observe center-of-mass motion of both droplets.} We assume the kinase is uniformly distributed in the simulation box, whereas the dephosphorylation occurs with a rate proportional to the local concentration of phosphatase. The combined reaction kinetics are $\partial\rho_\s/\partial t=-\partial\rho_\ns/\partial t=-k\rho_\s+(k'/\rho_\max)\rho_\p\rho_\ns$. We provide the full equations and simulation details in Appendix~\ref{sec:simulation-cond}. 

\begin{figure}
    \centering
    \includegraphics[width=0.99\linewidth]{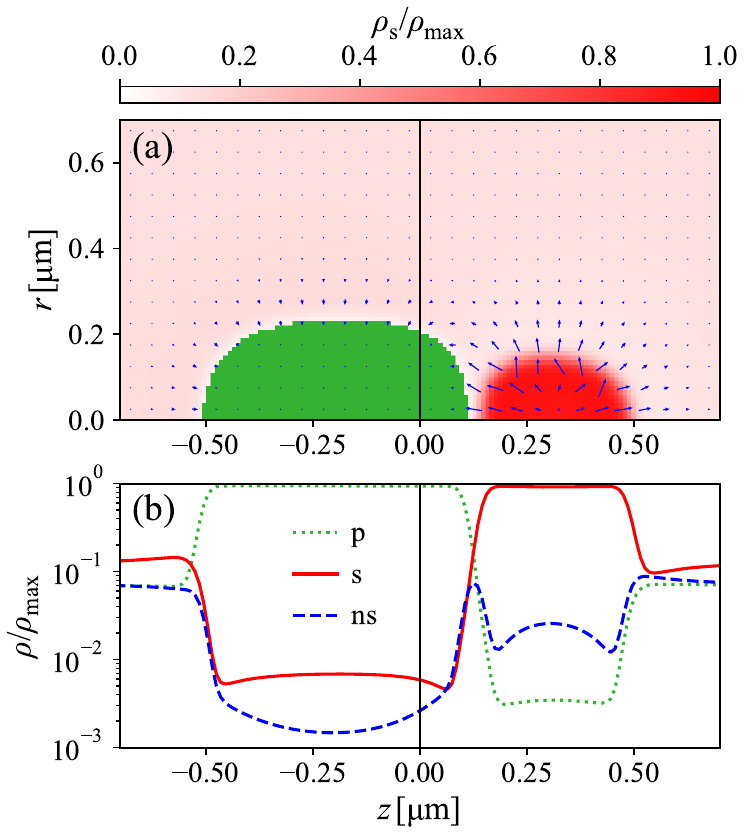}
    \caption{Cylindrically symmetric quasi-two-dimensional simulations of a self-propelling phosphatase- and sticky-condensate pair in a uniform solution of a kinase. (a) Steady-state concentration profile of the sticky species, $\rho_\s(r,z)$ (red, normalized by the maximal density, $\rho_\max=0.1\mathrm{nm}^{-3}$) along with a silhouette of the phosphatase condensate (green, in $(r,z)$ coordinates where $\rho_\p(r,z)\geq\rho_\max/2$). The ``pixels'' exactly represent our finite-volume density elements. Blue arrows show the sticky-protein flux $\vecJ_\s(r,z)$, computed on a $5\times5$-fold coarser grid (for clarity), where the arrow lengths are scaled such that the largest $|\vecJ|$ is five pixels ($0.05\mathrm{nm}$). (b) Line scan along the axis of symmetry ($r=0$) of the sticky, nonsticky, and phosphatase concentration profiles, $\rho_i(r=0,z)$ ($i=\s,\ns,\p$). Parameter values: average distance until phosphorylation, $\ell_k=0.40\micron$, and until dephosphorylation in a maximally dense phosphatase region, $\ell_k/5$, protein supersaturation relative to the spinodal-binodal difference $(\rho_{\s+\ns,\tot}-\rho_\bin)/(\rho_\spn-\rho_\bin)=0.9$; and total average phosphatase concentration $1.075\rho_\bin$. These plots are close-ups of a snapshot of a propelling state (drift velocity $1.6\micron/\s$); the total simulated-system size is $z\in(-1\micron,1\micron]$ and $r\in[0\micron,1\micron]$, with periodic and reflecting boundary conditions, respectively.}
    \label{fig:p-condensate}
\end{figure}

Indeed, scanning through phosphatase supersaturations and dephosphorylation rates (with fixed $\rho_\tot$ and $\ell_k$), we obtain a range of parameters for which the phosphatase condensate can maintain a single sticky droplet in its vicinity and self-propel. In Fig.~\ref{fig:p-condensate}, we plot the steady configuration that we find for one set of parameter values. This pair of condensates self-propel with a drift velocity of $1.6\micron/\s$. Since there is no direct repulsive interaction but only capillary opposition to deformations, this drift is smaller than that observed for the colloid. Nevertheless, this result supports the proposition made in Ref.~\cite{QiangPRL2025} that size-controlled droplets~\cite{ZwickerPRE15} can become mobile emulsions. 

Phenomena akin to active droplet self-propulsion have been reported~\cite{ZwickerROPP25}, \eg active droplets ``budding'' out of a macroscopic coexisting phase~\cite{QiangPRL2025}, droplet positions oscillating within confinement due to interactions with substrate molecules~\cite{DemarchiPRL2023}, and surface ``ripples'' of a large droplet exhibiting propulsion as capillary waves~\cite{RashoferPRR2025}. Here, we confirm that a single catalytic droplet of interest in isolation may be propelled by a sticky-protein droplet. This scenario therefore offers an experimental route to self-diffusiophoresis in micron-sized droplets, much smaller than the Marangoni swimmers discussed below.

\section{Discussion}\label{sec:discussion}

\subsection{Summary}

In this work, we explored the possibility of generating self-propulsion of an otherwise isotropic object in a homogeneous passive solvent, using active biomolecular condensates. Our proposed mechanism involves dissolving three components---a phase-separating (sticky) protein, a kinase that passivates the sticky protein, and a colloid-coating phosphatase that restores the protein to its sticky form. Despite the corresponding dynamics still being isotropic, homogeneous, and autonomous, we show via numerical simulations that a symmetry-breaking event can give rise to a steady traveling state. The combined action of the kinase in the bulk and the phosphatase on the surface of the sphere enables the nucleation, stabilization, and localization of a single droplet near the colloid surface. The condensate pushes on the colloid via direct repulsive interaction, and the resulting self-propulsion is persistent and exceptionally robust to off-trajectory perturbations, and achieves drift velocities exceeding a hundred colloid radii per second.

Our construction gives rise to several additional notable phenomena. First, as we increase the overall amount of sticky proteins by reducing kinase activity or increasing the total protein saturation, we observe steady states with multiple droplets. Taking sphere wetting as a point of reference (as the droplets are actively kept close to the colloid), the partially wet state---a single spherical cap---is the thermodynamically preferred state (being more stable even than the fully wet state~\cite{EralLangmuir11,SmithCS81}). Instead, the two- and six-droplet steady states we observe are thermodynamically unfavorable and are thus a direct consequence of activity. 

Specializing to the single-droplet state, we find that an immobilized colloid experiences unusually large net forces, exceeding $1\,\mathrm{nN}$. As each protein interacts with the colloid via entropic or electrostatic repulsion of order $\kt$~\cite{FleerBOOK98}, the imbalance between dilute- and dense-phase concentrations across a mesoscopic droplet accumulates to produce such large forces. However, upon freeing the colloid, the surrounding protein concentration profile must diffusively readjust as the colloid propels through the fluid: The dilute phase at the colloid's front must ``clear the way,'' whereas the dense droplet at the colloid's rear must ``keep up.'' At nanonewtons of force, the propulsion is limited by the rate of protein diffusion, so rapid colloidal drift is opposed. Indeed, once the steadily propelling state is reached, the net force subject to the reorganized (traveling) protein concentration profiles is smaller than that predicted by the immobile-colloid simulation by at least $3$ orders of magnitude. The fact that colloidal propulsion is limited by protein rearrangement means that perturbations misaligned with the direction of propulsion are subject to considerable resistance from the proteins. Indeed, the directed self-propulsion is highly resistant to Brownian motion as well as to large external driving forces due to a very stiff ($100\mathrm{pN}/\mathrm{nm}$) effective restoring force. Finally, owing to the rapid propulsion, the persistence length of colloidal trajectories is tens of body lengths.

We make a number of simplifying assumptions in our numerical formulation. For instance, the phosphatase must act sufficiently fast to be able to restore sticky proteins; otherwise, no droplet is able to form. In Appendix~\ref{sec:surface}, we confirm that the required phosphatase catalytic efficiencies are indeed within a physiological range. For completeness, in Appendix~\ref{sec:kinase} we discuss the assumptions underlying the kinase. In addition, we only consider parameter values that prevent sphere wetting, \eg through entropic or electrostatic repulsion of the proteins from the colloid~\cite{FleerBOOK98}. (A system consisting of oppositely charged colloids and proteins would require a separate study, wherein propulsion might be generated via a droplet pulling on the colloid by asymmetrically wetting it. A possible challenge to this scenario is the slow timescales associated with relaxation of wet states, possibly exhibiting severe multistability and hysteresis.) Furthermore, our model, consisting of a Flory-Huggins free energy for the proteins and an overdamped Langevin equation for the colloid, is rather simple and would likely require refinement for particular experimental systems. 

We ignored several hydrodynamic complexities: As the colloid moves through the surrounding fluid, it experiences the friction term of Eq.~\eqref{eq:drift} once steady Stokes flow is reached~\cite{LandauFluid1987}. However, in modeling the protein diffusion, we ignored the velocity profiles that surround a steadily moving sphere in a liquid~\cite{LandauFluid1987}; instead, we treated the proteins as if they diffused in an quiescent fluid. This is in fact a ``worst-case'' scenario for propulsion, as the proteins closer to the colloid would actually be transported along with the colloid via the no-slip boundary condition. Therefore, in experiment, we expect the proteins closer to the colloid to more readily keep up with the colloid leading to stronger propulsion forces and potentially additional deformation of the condensate. We have also ignored capillary effects such as Korteweg stress~\cite{AndersonARFM1998} or Marangoni flows~\cite{JambonPuilletNC24}, impacting the host-fluid and condensate flow profiles due to gradients in concentrations and local surface tension, respectively. Since the solute concentrations are low and the colloid and droplet are rather small, these effects should be negligible. For instance, $1\micron/\s$ droplet propulsion by Marangoni flows occurs for about $10\micron$ droplet~\cite{JambonPuilletNC24,IzriPRL14}; as this velocity scales with radius, for our droplets of around $0.1\micron$, the contribution to drift is expected to be very small. Furthermore, the simple Eq.~\eqref{eq:drift} must be adapted when considering complex and intracellular environments~\cite{StromCell2024}.  Overall, even if any of the above effects reduces the net force applied by the droplet, the force may decrease by up to $3$ orders of magnitude and protein diffusion will still be the limiting factor in the colloid's propulsion.

\subsection{Diffusiophoretic mechanisms}\label{sec:discussion-diff}

The self-propelled mechanism we found is a biologically realizable variant of self-diffusiophoresis. We begin with a brief overview of conventional diffusiophoresis~\cite{ShimCR2022}:\footnote{We are interested in bulk propulsion, so we ignore boundary effects and the emergent analogous phenomenon of diffusioosmosis.} Since colloids are often (negatively) charged in aqueous solutions, an electrostatic potential, called the zeta potential, forms at their surface. Thus, if an externally maintained salt-concentration gradient (or, more generally, any chemical-potential gradient~\cite{MarbachJFM2020,ShiPRF2025}) is present in the solution, the colloid would move parallel to that gradient~\cite{PrieveJFM1984,AndersonSPM1984,DeyNL15}. Importantly, diffusiophoresis was shown to affect cytoplasmic biomolecules of diverse sizes and functions~\cite{ShimNC2024,RammNP2021,SearPRL2019}. Moreover, biomolecular condensates were shown \textit{in vitro} to propel by diffusiophoresis given externally or other-condensate-imposed gradients via Marangoni flows~\cite{JambonPuilletNC24}. Note that the gradients need not be chemically induced---heating via light has been used to turn a near-critical solution to supercritical in the vicinity of a colloid, thereby creating a dense droplet that can push upon the colloid~\cite{ButtinoniJPCM2012}. More complex light-induced colloidal propulsion by pushing is possible, \eg using bacteria whose swimming velocity is dependent on light~\cite{PellicciottaNC23}. Since faster-swimming bacteria organize in lower-density regions, the high-density slow regions apply a higher overall active swim pressure on the colloid, thereby inducing propulsion. Interestingly, the ``solutes'' themselves are active agents in this case; in our work, activity is only involved indirectly, shaping the (passive) protein organization.

In this work, we set out to achieve self-propulsion, \ie colloidal motion in a homogeneous solution free of externally imposed gradients. Thus, we make the comparison to self-diffusiophoresis, wherein nonequilibrium processes occurring at the surface of the colloid give rise to an electrochemical-potential gradient in an otherwise homogeneous solution. Reassuringly for our purposes, recent advances in colloidal science allow the enzyme coating of colloids, which subsequently exhibit enhanced diffusion~\cite{MaNL15,DeyNL15}. Paradigmatic self-diffusiophoretic colloids are Janus particles, which are coated anisotropically with a catalyst that converts fuel molecules~\cite{MichelinJFM2014,HowsePRL02007,ThutupalliNJP2011,MaNL15}. Janus-like particles are still not quite our sought-after mechanism, as we require the colloid itself not to break rotational symmetry. 
One exemplary strategy for breaking symmetry of isotropic particles involves electrohydrodynamic coupling of a large population of so-called Quincke rollers~\cite{BricardNature2013}; the particles are insulating colloids that rotate spontaneously when immersed in a conducting fluid subject to an external electric field, thereby giving rise to rapid collective propulsion.
A potential advantage of propulsion along a symmetry-broken direction is that only an infinitesimal amount of energy or information is required to select the direction of motion. To our knowledge, there are only a handful of mechanisms for achieving self-diffusiophoresis of single isotropic catalytic objects by means of symmetry breaking, which we list below. 

One prevalent mechanism is spherically symmetric colloids isotopically emitting/absorbing solutes at a constant rate~\cite{MichelinPF2013}. Beyond a certain activity-related P\'{e}clet number, the isotropic fluid state is no longer stable---a small perturbation in solute concentration around the colloid gives rise to slip flows as particles are emitted/absorbed, which reinforces the asymmetry and thereby produces propulsion. This phenomenon has been experimentally realized for an isotropically coated colloid~\cite{ArcherAS23}, but mostly for isotropic liquid droplets~\cite{MichelinARFM2023,IzriPRL14,IzzetPRX2020,HokmabadPRX2021}. There, the instability and propulsion occur by virtue of active Marangoni flows. These Marangoni swimmers constitute a common experimental system used to study the emergence of isotropic-droplet self-propulsion.

Since the physical origin is different, we unsurprisingly observe several noteworthy, interrelated differences between Marangoni swimmers and our droplet-mediated colloid propulsion: (i) Self-diffusiophoresis of Marangoni swimmers is typically reported for $10$--$100\micron$-sized droplets, whereas the colloids we considered are not bigger than $1\micron$. (ii) The Marangoni drift-velocity magnitudes increase with droplet size as Marangoni flows become more prominent~\cite{MichelinARFM2023,JambonPuilletNC24}, whereas Fig.~\ref{fig:diagram_free} shows that our droplet-repulsion forces and thus the corresponding velocities are larger for smaller-radius droplets. (iii) The propulsion velocities of Marangoni swimmers are in the range $1$--$100\micron/\s$~\cite{MichelinARFM2023}, whereas our velocities are in the range $10$--$100\micron/\s$; the faster relative velocity (body lengths per second) in our case is likely due to the vast difference between the dilute- and dense-phase concentrations (whereas Marangoni swimmers typically explore an ``ideal'' dilute solution; the higher end of their velocity range indeed occurs for micellizing solutions~\cite{IzzetPRX2020}). (iv) Experimental realizations of Marangoni swimmers do not replenish the substrate (meaning that the propulsion is transient), yet they are very long lived (lasting for $10^3\s$)~\cite{IzriPRL14}. Without phosphatase, the droplets in our simulations dissolve within $10^{-3}\s$ because our droplets are a hundredfold smaller in radius (millionfold in volume). (v) Marangoni swimmers appear via a linear instability; since, to remain biologically realistic, we considered protein concentrations in the binodal regime, our droplets must form by nucleation, \eg via thermal noise~\cite{BinderROPP1987,ZiethenPRL2023}.

The few other autodiffusiophoretic mechanisms involve alternative types of couplings to hydrodynamic flow across scales. Nanometer-sized chemical motors, operating via thermophoretic coupling, have also been fabricated recently~\cite{deFigueiredoarXiv2026}. Another autodiffusiophoretic mechanism involves bubble generation and bursting via hydrogen peroxide breakdown, which also enables the propulsion of isotropic colloids~\cite{GregoryJPCC15}. [The above-mentioned light-tuned near-critical solution~\cite{ButtinoniJPCM2012} and bacterial bath~\cite{PellicciottaNC23} operate by a similar principle, wherein (molecular or agentic) particles that have been driven by light to accumulate on one side drive the propulsion of a colloid.] The last two mechanisms are more bioinspired: One example is based on particle aggregation, wherein particles, whose size is no less than a tenth of the colloid's, aggregate within a shell region (larger than the colloid) where the colloid is able to turn them into a ``sticky'' state~\cite{DecayeuxPRE21,DecayeuxEPJE22}. The aggregation is highly dynamical, meaning that the aggregates' interactions with the colloid push on it in random directions, thereby enhancing the colloid's diffusion. Another is based on collective actin treadmilling~\cite{BernheimGroswasserNature02,KawskaPNAS12}, in which actin polymerizes at the surface of a colloid to form a growing actin network. Continued polymerization behind the colloid generates stresses in the actin gel that propel the colloid forward while the network treadmills rearward. Upon introducing actin recycling, this propulsion mechanism can occur at cellular scales for very long durations~\cite{ColinEMBO23}.

Thus, we see that identifying novel mechanisms of self-diffusiophoresis for isotropic catalytic objects is still a developing field. Many of the above autodiffusiophoretic mechanisms may not be feasible in intracellular conditions, as they either may not scale down well to subcellular dimensions~\cite{MichelinARFM2023,JambonPuilletNC24,GregoryJPCC15} or they involve, \eg heating~\cite{ButtinoniJPCM2012,deFigueiredoarXiv2026} or fast turnover of gaseous matter~\cite{GregoryJPCC15}. It may not be surprising that biomolecular condensates permit a fast, stable, and persistent self-propulsion mechanism, as new desired functionalities of condensates are being uncovered for more than a decade. Having a dense phase is useful for mRNA sequestering (in, \eg stress granules~\cite{ShinSCI17}), reaction catalysis (\eg in the nucleolus~\cite{ShinSCI17} and in the carbon-concentrating pyrenoid~\cite{HeNCB2026}), and lowering noise (\eg during gene expression~\cite{KlosinSci2020}). More pertinent to our work, condensed protein phases are implicated in shaping cellular structures through force generation and other mechanical properties~\cite{StromCell2024,SemeigazinARXIV25}. Thus, going forward, it is an encouraging notion that biocondensates may autonomously push on catalytic objects via a simple mechanism. 

\subsection{Statistical thermodynamics of self-propulsion}\label{sec:discussion-thermo}

From a theoretical perspective, in generating spontaneous self-propulsion, the system must be consuming energy at a steady rate~\cite{PelitiBOOK2021}. This is because we explicitly break detailed balance: Indeed, the purpose of the phosphorylation cycle (involving an ATP-consuming kinase) is to actively maintain a protein-rich droplet in the vicinity of the colloid, which would otherwise repel the protein. Qualitatively, this cycle gives rise to three protein states---dense-phase sticky (``$\mathrm{den}$''), dilute-phase sticky (``$\mathrm{dil}$''), and inert (``$\ns$''). They interconvert via (thermodynamics-driven) phase separation ($\mathrm{den}\rightleftharpoons\mathrm{dil}$), (energy-consuming kinase-driven) passivation ($\mathrm{den}\to\ns$ and $\mathrm{dil}\to\ns$) and (non-energy-consuming phosphatase-driven) depassivation ($\ns\to\mathrm{dil}$). Since breaking detailed balance requires at least three states, we therefore, by construction, formulated a minimal scheme, involving exactly three states and three independent processes switching between these states.\footnote{Note that indeed the previously mentioned actin-treadmilling- and aggregation-based autodiffusiophoretic mechanisms~\cite{ColinEMBO23,DecayeuxPRE21} also involve three effective protein states (\eg active actin free monomer, active fiberous actin, and passive actin free monomer). In contrast, the Marangoni swimmers require only two particle states~\cite{MichelinPF2013}---absorbed and free particle. The latter system, however, is not in a steady state: The solute sustaining the symmetry-broken state is progressively depleted. This distinction is analogous to that between Refs.~\cite{BernheimGroswasserNature02,ColinEMBO23}: While the former draws from a (large) finite actin pool, the latter introduces ATP-driven recycling through a third effective actin state. An analogous recycling step in Ref.~\cite{MichelinPF2013} would likewise render its mechanism comparable to the three-state cycle considered here.} This approach results in a constant cycling of protein through the qualitative $\mathrm{dil}$--$\mathrm{den}$--$\ns$ phase space at steady state. Generating self-propulsion is then a matter of coupling this breaking of detailed balance into mechanical motion. We have done so via protein-colloid repulsion or capillary exclusion of the phosphatase and sticky-protein condensate. 

We note that our mechanism is related to systems involving particles~\cite{AgudoCnalejoPRL19} and fields~\cite{SahaPRX20,AvniPRL2025} that interact nonreciprocally. Models of such systems reveal emergent collective phenomena, \eg self-propelled bands originating from a linear instability~\cite{SahaPRX20}. In comparison, the excitations we found occur via nucleation rather than instability~\cite{BinderROPP1987}, and are localized to compact regions (sticky-rich droplets). Furthermore, while these nonreciprocal models implicitly assume that some suitable energy-consuming microscopic interactions give rise to the nonreciprocal interactions, our system is very explicit in its components and its (microscopically reciprocal) interactions. In fact, our system may therefore serve as one microscopic model for an emergent nonreciprocal interaction among two particle types by viewing the droplet as a particle: The colloid is repelled from the droplet, but the droplet appears as if attracted to the colloid, and hence follows the colloid. The interaction between several such self-propelled catalytic objects may give rise to collective phenomena that are worthy of further study. (For instance, previous works on active-aggregation-mediated~\cite{DecayeuxSM23} and actin-mediated~\cite{GuerinCB25} propulsion exhibit rich competition dynamics over resources.)

As we have discussed in Sec.~\ref{sec:discussion-diff}, it is already nontrivial to obtain self-propulsion in an isotropic, homogeneous, and autonomous system via symmetry breaking. Yet, a fascinating next step would be to construct this means of self-propulsion in an energetically efficient manner. For reference, kinesin---a paradigmatic transport protein---has an efficiency which is estimated at $20\%$~\cite{ArigaPRL18}. Is it possible to construct a microswimmer more energetically efficient than biological motor proteins?

Here, the immobile colloid experienced forces that are $3$ orders of magnitude larger than the mobile one. Clearly, the vast majority of the energy is being dissipated through viscous resistance of the fluid to the rearranging proteins. Therefore, we cannot expect good energetic efficiency: Assuming a single ATP consumed per one passivation event, the energy per second consumed by the kinase is around $10^{11}\kt/\s$, whereas the viscous dissipation due to the colloid's motion is about $10^4\kt/\s$ (see Appendix~\ref{sec:surface-cat}). Indeed, merely $10^{-7}$ of the consumed energy is being utilized for propulsion. 

Coincidentally, if the propulsion is quasistatic compared to the proteins' diffusive rearrangement and the forces could indeed reach up to $1\,\mathrm{nN}$ (\eg for a colloid burdened by a large cargo), the dissipation would increase to $\mu_\col F^2\sim10^{11}k_\mathrm{B}T/\s$. This case suggests that our mechanism has the potential of propelling with $\mathcal{O}(1)$ efficiency. At the same time, our model is agnostic to the energetic cost entailed by each passivation; it may have been one or more ATPs or an entirely different fuel molecule. At no point did we impose the second law of thermodynamics to forbid the viscous dissipation by the colloid to exceed the energy input to the kinase. Thus, it could be insightful to construct a thermodynamically consistent model that correctly accounts for the energetic costs~\cite{MarkovichPRX2021}. One possible route is to begin with a fully reversible theory, wherein both the phase-separation and interconversion-reactions kinetics are derived from a total free energy (Eq.~\eqref{eq:CH}). One could then gradually increase a controllable chemical-potential bias (``fuel'') favoring the phosphorylation reaction, until a motility threshold is crossed.

\subsection{Conclusion}

To conclude, active biomolecular condensates permit biological rapid and persistent motion of a micron-sized sphere via symmetry breaking in an otherwise homogeneous, isotropic, and autonomous environment. We expect our design principles---keeping a droplet close to the object of interest by passivation-activation cycles---to hold in general, and therefore to be experimentally realizable \textit{in vitro} for arbitrary catalytic spheres---coated colloid, enzymatic condensates, and other biological structures. We do not rely on motor proteins; instead, we use a minimal set of components with simple interactions and functions. It would be particularly interesting to observe \textit{in vivo} an active droplet pushing an object along the entire length of a cell.

\acknowledgements
We thank Eric Dufresne, Janni Harju, Howard Stone, and Jared Toettcher for insightful discussions. B.S. acknowledges support from the Princeton Center for Theoretical Science. This work was supported in part by Princeton University through the Center for the Physics of Biological Function, and was performed in part at Aspen Center for Physics, which is supported by National Science Foundation Grant No. PHY-1607611.

\appendix

\section{Simulation details}\label{sec:simulation}

\subsection{Parameter values}\label{sec:simulation-para}

\begin{table*}
\begin{tblr}{cccc}
        Parameter & Symbol & Value (main text) & Value (confirmation) \\
        \hline\hline
        Molecular diffusion constant & $\mu_\mol\kt$ & $220\micron^2/\s$ & N/A \\
        Dense-dilute interface width & $\xi$ & $10\mathrm{nm}$ & N/A\\
        \hline
        Average distance until phosphorylation & $\ell_k=(\mu_\mol\kt/k)^{1/2}$ & See main text figures & \\
        Relative protein supersaturation & $(\rho_\tot-\rho_\bin)/(\rho_\spn-\rho_\bin)$ & See main text figures & \\
        Colloid radius & $R$ & See main text figures & \\
        \hline
        Maximal protein concentration & $\rho_\max$ & $10^2\xi^{-3}$ & \\
        Flory-Huggins parameter & $\chi$ & $3$ & $3.2$ and $3.5$ \\
        Protein-colloid repulsion magnitude & $U_0/\kt$ & $3$ & $4$ and $5$ \\
        Protein-colloid repulsion range & $\ell_U$ & $\xi$ & $0.8\xi$ and $2\xi$ \\
        \hline
        Spatial discretization & $\Delta x/\xi$, $\Delta y/\xi$, and $\Delta z/\xi$ (and $\Delta r/\xi$) & $1$ & $0.8$ \\
        Temporal discretization & $\mu_\mol\kt\Delta t/\xi^2$ & $0.002$ & $0.001$ and $0.004$
    \end{tblr}
    \caption{Parameter values. The first group of parameters (molecular diffusion constant and dense-dilute interface widths) are the parameters we set to $1$ in our simulations; units are reintroduced back by rescaling all length and timescales by $\xi$ and $\xi^2/(\mu_\mol\kt)$, respectively. The second group of parameters (average distance until phosphorylation through colloid radius) includes the parameters we change systematically in the main text. The remaining parameters are fixed throughout the main text at the values indicated in the ``Value (main text)'' column. We have confirmed that our phase diagram (Fig.~\ref{fig:diagram_static}) remains qualitatively unchanged for simulations with parameter values indicated in the ``Value (confirmation)'' column.}
    \label{tab:params}
\end{table*}

We first specify the parameter values, summarized in Table~\ref{tab:params}. Without loss of generality, we may measure length in units of the dense phase-dilute phase interface width $\xi$ and time in units of molecular diffusion $\xi^2/(\mu_\mol\kt)$. Nonetheless, we will assign particular values to these quantities to make concrete magnitude estimates of the typical forces and propulsion velocities. To avoid being overly specific, we assign parameter values from expected orders of magnitude as follows: A typical size of a small protein or a polymer is about $1\mathrm{nm}$, such that we may expect a condensate interface width of order $\xi=10\mathrm{nm}$~\cite{FleerBOOK98}. From the Stokes-Einstein formula in water at room temperature, the diffusion constant of a nanometer-scale molecule is about $\mu_\mol\kt\simeq220\micron^2/\s$. We take the maximal permitted density to be a tenth of the maximally packed protein density, so $\rho_\max=0.1\mathrm{nm}^{-3}$. For phase separation to occur, we choose $\chi=3$ (corresponding to a partition coefficient of about $13$, which is typical~\cite{VarmaARXIV26})). Simulations using higher $\chi$ values ($3.2$ and $3.5$) produce qualitatively similar phase diagrams (not shown).

As for the colloid-protein repulsion,\footnote{We also performed simulations with $U_0=0$ and $U_0=\kt$, which permitted wetting. However, the phase diagrams were significantly more intricate to control due to the exceedingly long running times needed to reach stable wetting morphologies. Therefore, we restricted the current study to a nonwetting regime.} we choose, for simplicity, an exponentially decaying form,
\begin{equation}
    U(r)=U_0e^{-(r-R)/\ell_U}.
\end{equation}
We take $U_0=3\kt$, which is the lowest integer multiple of $\kt$ that prevents wetting. Typical electrostatic (\eg double-layer) repulsions and entropic repulsions are of order of a few $\kt$s~\cite{IsraelachviliBOOK11}. We choose $\ell_U=\xi=10\mathrm{nm}$~\cite{FleerBOOK98}, as this repulsion arises from similar interactions to those underlying the dense-dilute interface width. We confirm that small variations in $U_0$ (to $4\kt$ or $5\kt$) and $\ell_U$ (to $8\mathrm{nm}$ or $20\mathrm{nm}$) do not prevent self-propulsion; the propulsive band in our phase diagrams simply moves and changes shape in the $\ell_k$--$\rho_\tot$ plane. 

In Appendix~\ref{sec:surface} we justify the fast nonsticky-to-sticky conversion limit. Ultimately, it requires the catalytic efficiency of the phosphatase $k_\mathrm{cat}/K_\mathrm{m}$ to lie above $10^5\mathrm{M}^{-1}\s^{-1}$. This requirement is quite reasonable given that the typical enzyme in nature has this efficiency~\cite{BarEvenBC11}, and phosphatases are generally much faster, with efficiency often exceeding $10^7\mathrm{M}^{-1}\s^{-1}$~\cite{SimopoulosBC94,BarEvenBC11}. Using the boundary condition explained there, finite reaction rates (with a reaction length $\ell_\gamma=10\mathrm{nm}$) on the interface also result in qualitatively similar phase diagrams; the propulsive band simply shifts to larger $\ell_k$ and $\rho_\tot$. 

\subsection{Simulation: Self-propelled colloid}\label{sec:simulation-coll}

We perform finite-volume simulations of Eqs.~\eqref{eq:PDE_s} and~\eqref{eq:PDE_ns}, where the fluxes are explicitly given by
\begin{widetext}
\begin{eqnarray}
    -\frac{\vecJ_\s(\vecr,t)}{\mu_\mol\kt}&=&\biggl[\frac{\rho_\max-\rho_\ns(\vecr,t)}{\rho_\max-\rho_\s(\vecr,t)-\rho_\ns(\vecr,t)}-\frac{2\chi}{\rho_\max}\rho_\s(\vecr,t)\biggr]\grad\rho_\s(\vecr,t)+\frac{\rho_\s(\vecr,t)}{\rho_\max-\rho_\s(\vecr,t)-\rho_\ns(\vecr,t)}\grad\rho_\ns(\vecr,t)\nonumber\\&&-\frac{2\chi\xi^2}{\rho_\max}\rho_\s(\vecr,t)\grad\nabla^2\rho_\s(\vecr,t)+\rho_\s(\vecr,t)\grad U(|\vecr|),\label{eq:flux_expl_s}\\
    -\frac{\vecJ_\ns(\vecr,t)}{\mu_\mol\kt}&=&\frac{\rho_\ns(\vecr,t)}{\rho_\max-\rho_\s(\vecr,t)-\rho_\ns(\vecr,t)}\grad\rho_\s(\vecr,t)+\frac{\rho_\max-\rho_\s(\vecr,t)}{\rho_\max-\rho_\s(\vecr,t)-\rho_\ns(\vecr,t)}\grad\rho_\ns(\vecr,t)+\rho_\ns(\vecr,t)\grad U(|\vecr|).\label{eq:flux_expl_ns}
\end{eqnarray}
\end{widetext}
The densities are computed at the center of each volume element and the fluxes are computed at the corresponding faces of each cell. Time propagation is performed using the forward-Euler method. 

Since $\rho_\tot$ lies within the binodal region of the sticky-protein phase diagram, the system is metastable rather than unstable~\cite{BinderROPP1987}. (In other words, a dense phase will not appear spontaneously by spinodal decomposition.) Instead, droplets may only form via a nucleation process, requiring a finite-size perturbation. We detail this point in Appendix~\ref{sec:nucl}. For purposes of systematic study, throughout all simulations, our initial conditions consist of a single spherical droplet of radius $10\xi$ placed at a separation $5\xi$ between its surface and that of the colloid. In this way, we introduce the broken symmetry into our deterministic equations. Since experiments are necessarily performed at finite temperatures, nucleation is guaranteed to occur.

All simulation results in the main text are for a three-dimensional cubic lattice. The system size is $480\xi\times480\xi\times480\xi$. To efficiently simulate these sizes in a reasonable time, we use the JAX software~\cite{jax2018github}, and split the system into two grids: The first is a fine grid with temporal discretization $\Delta t=0.002 \xi^2/(\mu_\mol\kt)$ and spatial discretization $\Delta x=\Delta y=\Delta z=\xi$. It occupies the cube $|x|,|y|,|z|<36\xi+R$, where the colloid of radius $R$ is placed at the origin, $x=y=z=0$. It is sufficiently large that the droplets never grow toward the finer grid's boundary. (In Fig.~\ref{fig:force_profile}, we only show cross sections of this inner fine grid.) The colloid is represented on the grid as all points for which $x^2+y^2+z^2\leq R^2$ [see ``pixelation'' in Figs.~\ref{fig:force_profile}(a) and~\ref{fig:force_profile}(b)]. To enforce the boundary conditions at the colloid's surface [Eqs.~\eqref{eq:BCinstantconv} and~\eqref{eq:BCnoflux}], we set both sticky and nonsticky fluxes perpendicular to its faces to zero at each time step and convert all nonsticky species in the nearest-neighbor volume elements (excluding diagonals) to sticky. 

The first, inner fine grid is surrounded by the second outer, coarser grid: $\Delta t=0.2\xi^2/(\mu_\mol\kt)$ and $\Delta x=\Delta y=\Delta z=8\xi$. For this grid, we can increase the discretization in time and space as no droplets can form there, and the phosphatase operates on length scales $\ell_k\gg8\xi$. We ``stitch'' the two grids together by linearly interpolating the density field at their boundaries and accumulating the fine-grid surface fluxes prior to updating the coarse grid. We use periodic boundary conditions for the overall system (whose edges only involve the coarse cells). Each simulation runs until $8\times 10^4\xi^2/(\mu_\mol\kt)$, by which point the system attains a steady state: These times are not long compared with those required to observe sustained colloidal propulsion: For $a\sim1\micron$ and the observed $v\sim10-100\micron/\s$, traversing one colloid radius requires $a/v\sim10-100\mathrm{ms}$. Our simulations, however, are not intended to resolve long-time propulsion, which is addressed by the dimensional analysis in Sec.~\ref{sec:Bro}. Rather, they assess the stability of the condensate-colloid pair in the corotating--cotranslating frame. The simulated time windows are therefore chosen such that the force and drift velocity converge monotonically to within $1\%$ of their final values (and typically reach $0.01\%$), demonstrating that the relevant steady state is reached.

Unless specified otherwise, the additional ``sanity-check'' simulations are performed for the immobile-colloid cases with $R/\xi=15,20,30$. To save running time, we use a quasi-two-dimensional geometry on a cylinder of dimensions $200\xi$ in $z$ by $100\xi$ in $r$, whose central axis passes through the center of the colloid and the initial droplet. This simplified approach is motivated by the cylindrical symmetry of the no-, single-, and two-droplet configurations (along the axis connecting the droplets and the colloid). As a result, the six-droplet configuration manifests instead as two-droplet-and-disc configurations. The temporal discretization is $\Delta t=0.004 \xi^2/(\mu\kt)$ and the spatial one is $\Delta z=\Delta r=\xi$ everywhere. [Repeating the simulations for $\Delta t=0.001\xi^2/(\mu_\mol\kt)$ and $\Delta z=\Delta r=0.8\xi$ reveals only minor quantitative changes in steady colloid drift.] Furthermore, the propulsive band is wider than that of Fig.~\ref{fig:diagram_static} at larger $\ell_k$, due to a finite-size effect. Increasing the system size to $400\xi$ in $z$ and $200\xi$ in $r$ yields a phase diagram almost identical to that of the main text. We use periodic boundary conditions along $z$ and reflecting boundary conditions at $r$. Each simulation runs until $4\times 10^4\xi^2/(\mu_\mol\kt)$, which attains a steady state.

\subsection{Passivation kinetics}\label{sec:kinase}

We consider average distance to passivation within the range $\ell_k\in[32\xi,72\xi]$. We choose this range of values to avoid finite-size effects: The systems we simulate are $480\xi$ in the full three-dimensional simulations, and $200\xi$ in the quasi-two-dimensional ones. These choices avoid early truncation of the exponential decay in concentration of sticky particles away from the colloid. Since the passivation rate is inversely proportional to that distance squared, $k=\mu_\mol\kt/\ell_k^2$ (see Sec.~\ref{sec:PDEs}), this case in fact amounts to a scan of about an order of magnitude in rate, $k\in[2150\s^{-1},420\s^{-1}]$, which may support the robustness of our mechanism.

However, we note that with a concentration of about $1\mathrm{mM}$ of dissolved enzyme, the above range of rates may require an efficient passivator, $k_\mathrm{cat}/K_\mathrm{m}\sim10^5\mathrm{M}^{-1}\s^{-1}$ (\eg a urease~\cite{TestaNC21}, but not necessarily a kinase). To confirm that our propulsion mechanism may take place with a lower passivator concentration or efficiency, we run several immobilized-colloid simulations to see if our mechanism extends to larger $\ell_k$ (smaller $k$). First, as a benchmark, quasi-two-dimensional simulations for $R=20\xi$ in a system threefold larger in length ($600\xi$ along $z$ and $300\xi$ along $r$) for $\ell_k/\xi\in\{150,200, 225, 250, 275\}$ and $(\rho_\tot-\rho_\bin)/(\rho_\spn-\rho_\bin)\in\{0.1,0.15,0.2,0.25,0.3\}$ showed, in all cases, that an initial droplet has dissolved by $8\times10^4\xi^2/(\mu_\mol\kt)$, confirming that a slow kinase may oppose a fast phosphatase. 

\begin{figure}
    \centering
    \includegraphics[width=0.99\linewidth]{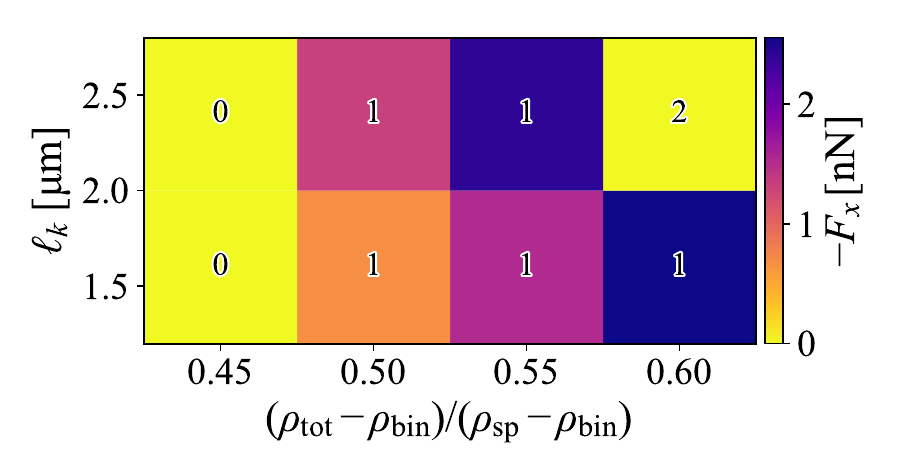}
    \caption{Phase diagrams for three-dimensional simulations of an immobilized colloid with radius $R=0.2\micron$, with significantly slower kinase (larger $\ell_k$) compared to Fig.~\ref{fig:diagram_static}(a) and concentration $\rho_\tot$ not significantly lower.}
    \label{fig:slowkinase}
\end{figure}

However, to confirm that a droplet can still be stabilized, since the passivator is substantially slower than in previous simulations, we have to run much longer simulations [note that $\ell_k^2/(\mu_\mol\kt)$ is on the order of the previous simulation durations $8\times10^4\xi^2/(\mu_\mol\kt)$]. We likewise want to eliminate finite-size effects. Our three-dimensional JAX-based simulations allow us to scale up the simulations further: We increase the inner finer grid to occupy a larger cube $|x|,|y|,|z|<48\xi+R$ (as the seeded droplet transiently coarsens to a larger size than in the previous, smaller-$\ell_k$ simulations by the time the slower kinase here begins reducing the droplet's size), the overall system size to $960\xi\times960\xi\times960\xi$ (to accommodate the larger distance to phosphorylation, $\ell_k$), and the total runtime to $4\times10^5\xi^2/(\mu_\mol\kt)$ (to accommodate the larger time between phosphorylations, $\ell_k^2/(\mu_\mol\kt)$). With these heavily demanding simulation dimensions, we obtain the results presented in Fig.~\ref{fig:slowkinase}. Reassuringly, our mechanism persists for substantially slower kinase activity: for these two $\ell_k$s, $k=\{86\s^{-1},38\s^{-1}\}$, and $k_\mathrm{cat}/K_\mathrm{m}\sim10^4\mathrm{M}^{-1}\s^{-1}$. Furthermore, the forces appear to be larger than in Fig.~\ref{fig:diagram_static}(a). Since $(\rho_\tot-\rho_\bin)/(\rho_\spn-\rho_\bin)\geq0.45$, there is still a considerable range of $\rho_\tot$ until $\rho_\bin$, which allows us to increase $\ell_k$ even further.

Another point worth considering is that the passivating enzymes and/or their cofactors may be excluded from or slowed down in the dense phase~\cite{DaiNRB23,TestaNC21} (though not always~\cite{HeNCB2026}). This method would slow down the passivation and breakdown of droplets. We expect such a slowdown to increase the stable droplet size~\cite{ZwickerPRE15}. To confirm, we run quasi-two-dimensional simulations for $R=20\xi$ with the same range of $\ell_k$s and $\rho_\tot$s as in Fig.~\ref{fig:diagram_static}(a), but replacing $\pm k\rho_\s(\vecr,t)$ in Eqs.~\eqref{eq:PDE_s} and~\eqref{eq:PDE_ns} with $\pm k[1-\rho_s(\vecr,t)-\rho_\ns(\vecr,t)]\rho_\s(\vecr,t)$, thereby imitating free-volume exclusion of the kinase. We find that our expectation holds---for the same $\ell_k$ and $\rho_\tot$, the droplets grow in size, as do the repulsion forces (the forces become as large as $7\mathrm{nN}$), and overall compared to Fig.~\ref{fig:diagram_static}(a) we find a notably wider single-droplet region in $\ell_k$--$\rho_\tot$ space [which has grown to the size of the single-droplet region in Fig.~\ref{fig:diagram_static}(b)].

\subsection{Consequences of nonfluctuating phase-field model}\label{sec:nucl}

Throughout this work, the only fluctuations we considered were Brownian noise in the colloidal dynamics, Eq.~\eqref{eq:drift}. Namely, Eqs.~\eqref{eq:PDE_s} and~\eqref{eq:PDE_ns} for the protein concentrations do not contain thermal noise. A full, fluctuating phase-field model would include an additional conservative noise, amending Eq.~\eqref{eq:flux} to
\begin{multline}
    \vecJ_i(\vecr,t)=-\mu_\mol\rho_i(\vecr,t)\grad\frac{\delta G}{\delta\rho_i(\vecr,t)}\\+\sqrt{2\kt\mu_\mol\rho_i(\vecr,t)}\boldsymbol\zeta_\rho(\vecr,t)\label{eq:flux_noisy}
\end{multline}
where $\boldsymbol\zeta_\rho$ satisfies the fluctuation-dissipation theorem, $\langle\boldsymbol\zeta_\rho(\vecr,t)\boldsymbol\zeta_\rho(\vecr',t')\rangle=\boldsymbol1\delta(\vecr-\vecr')\delta(t-t')$~\cite{deGrootBOOK84,ZiethenPRL2023}. Simulating fluctuating fields via Eqs.~\eqref{eq:PDE_s} and~\eqref{eq:PDE_ns} with the noisy fluxes of Eq.~\eqref{eq:flux_noisy} would be exceptionally challenging. Ensuring the stability of fluctuating-hydrodynamics simulations is still a largely unresolved numerical issue, requiring careful regularization techniques~\cite{JinPRE2026}. In short, the Courant--Friedrichs--Lewy condition, $\Delta t<\Delta x^4$, is in conflict with the diffusion term being able to smooth out the noise term $\Delta t^{1/2} >\Delta x^{-1/2}$, leading to an inherently unstable simulation unless the noise amplitude is restricted to be unphysically small. Even if this issue were to be resolved, from a practical perspective the selection of random noise in each time step would slow down our already demanding simulations by a considerable factor. Thus, we defer the treatment of fluctuating fields to future works. Yet, omitting field fluctuations means that our simulations may not capture two effects, which we will thus consider qualitatively in the following.

The first effect is our model's inability to capture the diffusion of the droplet around the colloid, considered in Sec.~\ref{sec:Bro}. As the concentration fields fluctuate, one may expect the dense-phase droplet to vary stochastically in time. Despite the presence of noise, however, it is known that, once a droplet has nucleated and grown to a mesoscopic size, noise only contributes small surface fluctuations~\cite{BinderROPP1987} and overall diffusion~\cite{BrayAP94} without destabilizing the droplet. Importantly, this fact has been shown repeatedly for varied classes of active droplets as well, such as the active model B+~\cite{FaustiPRL21,CatesPRL23} and active Brownian particles~\cite{LangfordPRE2024,LangfordJCP2025}. Thus, we can safely regard the droplets as rigidlike objects, exhibiting small surface fluctuations and diffusing in space. As we have shown in Sec.~\ref{sec:Bro}, the droplet and colloid remain at steady relative positions, implying that the pair is a practically rigid ``dumbbell.'' Thus, the diffusion of the droplet and colloid manifests only through a diffusion of the dumbbell's center of mass and of its orientation. Instead of running the above-mentioned, very costly fluctuating field-theory simulations, we make order-of-magnitude estimations in Sec.~\ref{sec:Bro}. There, we showed that spatial diffusion is small relative to the propulsion and that the reorientation, although not prohibited, occurs over a large persistence length of tens of colloidal radii.

The second effect our deterministic field model cannot capture is the nucleation of droplets. Since the sticky concentrations are within the binodal region, we should not expect droplets to nucleate spontaneously in the absence of noise. According to classical nucleation theory, thermal fluctuations enable the nucleation of droplets by overcoming the energetic barrier for nucleation~\cite{BinderROPP1987}. Particularly promising is the fact that droplet nucleation indeed occurs in the aforementioned active phase-separating models~\cite{CatesPRL23,LangfordJCP2025} but, most crucially for us, also in chemically active emulsions~\cite{ZiethenPRL2023,ChoPRL2023}.

As we cannot simulate fluctuating fields explicitly to permit the nucleation of droplets, we instead provide a plausibility argument as follows: We perform three-dimensional, noiseless simulations of an immobile $R=0.2\micron$ colloid with initial concentrations $\rho_\s(\vecr,0)$ and $\rho_\ns(\vecr,0)$ corresponding to the equilibrium (droplet-free) $\chi=0$ case. We inspect the fate of the system upon quenching it back to $\chi=3$. We consider $(\rho_\tot-\rho_\bin)/(\rho_\spn-\rho_\bin)=0.85$ and scan through $\ell_k=0.48\micron,0.56\micron,0.64\micron,0.72\micron$. Interestingly, $\ell_k=0.72\micron$ is able to bring forth six droplets, whereas the other $\ell_k$s remain droplet-free. Note that $\ell_k=0.72\micron$ is deep within the six-steady-droplets region of Fig.~\ref{fig:diagram_static}. For slow enough kinase, even though $\rho_\tot<\rho_\spn$, the repulsive colloid combined with $\chi>0$ concentrates sticky particles together in ``aura'' around the colloid sufficiently to exceed the spinodal concentration $\rho_\spn$ and nucleate the six droplets, which are stable in this regime. Thus, despite the below-spinodal concentration, droplets readily form without noise for some parameter values. This finding suggests that the nucleation barrier is effectively reduced compared to passive droplets, due to what we interpret as a density-driven heterogeneous nucleation. Since passive homogeneous nucleation is guaranteed to occur in all experiments, it should indeed be the case that this facilitated heterogeneous nucleation will also occur.

\subsection{Simulation: Self-propelled condensate}\label{sec:simulation-cond}

To extend the generality of our propulsion scenario, we perform equivalent simulations to Sec.~\ref{sec:simulation-coll}, but with a ``phosphatase condensate'' (p-condensate) replacing the colloid. We again use the quasi-two-dimensional (cylindrical) geometry, where the p-condensate has an initial radius $20\xi$ and is centered at the origin, while a sticky-protein condensate (droplet), of initial radius $10\xi$, is placed with a $5\xi$ gap between the surfaces. The system size is $200\xi$ in $z$ with periodic boundary conditions and $100\xi$ in $r$ with reflecting boundary conditions. Since the phosphatase condensate is penetrable by the sticky and nonsticky proteins and solvent (subject to entropic repulsion), there are no additional boundary conditions. The discretizations are $\Delta t=0.004\xi^2/(\mu_\mol\kt)$ and $\Delta z=\Delta r=\xi$. (The nonperfectly circular shape of the condensates in Fig.~\ref{fig:p-condensate} can be improved by refining the spatial discretization.) Since coarsening is a very slow process, we run the simulations for longer times, $5\times10^7\Delta t$.

The model is as follows: We denote the phosphatase density field as $\rho_\p$. The bulk Flory-Huggins free energy of Eq.~\eqref{eq:FH} changes to
\begin{multline}
    \frac{g(\rho_\s,\rho_\ns,\rho_\p)}{\kt}=\rho_\s\ln\rho_\s+\rho_\ns\ln\rho_\ns+\rho_\p\ln \rho_\p\\+(\rho_\max-\rho_\s-\rho_\ns-\rho_\p)\ln(\rho_\max-\rho_\s-\rho_\ns-\rho_\p)\\-\frac{\chi}{\rho_\max}\rho_\s^2-\frac{\chi}{\rho_\max}\rho_\p^2,\label{eq:FHp}
\end{multline}
where, for simplicity, we use the same $\chi=3$ for both the sticky protein and phosphatase. This approach favors self-interactions and thus (orthogonal) phase separation. Likewise, the Cahn-Hilliard free-energy functional of Eq.~\eqref{eq:CH} becomes
\begin{multline}
    \frac{G[\rho_\s,\rho_\ns,\rho_\p]}{\kt}=\int \d\vecr \biggl\{\frac{g(\rho_\s(\vecr),\rho_\ns(\vecr),\rho_\p(\vecr))}{\kt}\\+\frac{\chi\xi^2}{\rho_\max}|\grad\rho_\s(\vecr)|^2+\frac{\chi\xi^2}{\rho_\max}|\grad\rho_\p(\vecr)|^2\biggr\},\label{eq:CHp}
\end{multline}
where the phosphatase interface term replaces the colloidal surface repulsion---the p-condensate and droplet effectively repel only due to each one's preference for self interaction.

All species follow a flux given by Eq.~\eqref{eq:flux} (with $i=\s,\ns,\p$); explicitly,
\begin{widetext}
\begin{eqnarray}
    -\frac{\vecJ_\s(\vecr,t)}{\mu_\mol\kt}&=&\biggl[\frac{\rho_\max-\rho_\ns(\vecr,t)-\rho_\p(\vecr,t)}{\rho_\max-\rho_\s(\vecr,t)-\rho_\ns(\vecr,t)-\rho_\p(\vecr,t)}-\frac{2\chi}{\rho_\max}\rho_\s(\vecr,t)\biggr]\grad\rho_\s(\vecr,t)\nonumber\\&&+\frac{\rho_\s(\vecr,t)}{\rho_\max-\rho_\s(\vecr,t)-\rho_\ns(\vecr,t)-\rho_\p(\vecr,t)}\grad[\rho_\ns(\vecr,t)+\rho_\p(\vecr,t)]-\frac{2\chi\xi^2}{\rho_\max}\rho_\s(\vecr,t)\grad\nabla^2\rho_\s(\vecr,t),\label{eq:flux_p_s}\\
    -\frac{\vecJ_\ns(\vecr,t)}{\mu_\mol\kt}&=&\frac{\rho_\max-\rho_\s(\vecr,t)-\rho_\p(\vecr,t)}{\rho_\max-\rho_\s(\vecr,t)-\rho_\ns(\vecr,t)-\rho_\p(\vecr,t)}\grad\rho_\ns(\vecr,t)\nonumber\\&&+\frac{\rho_\ns(\vecr,t)}{\rho_\max-\rho_\s(\vecr,t)-\rho_\ns(\vecr,t)-\rho_\p(\vecr,t)}\grad[\rho_\s(\vecr,t)+\rho_\p(\vecr,t)],\label{eq:flux_p_ns}\\
    -\frac{\vecJ_\p(\vecr,t)}{\mu_\mol\kt}&=&\biggl[\frac{\rho_\max-\rho_\s(\vecr,t)-\rho_\ns(\vecr,t)}{\rho_\max-\rho_\s(\vecr,t)-\rho_\ns(\vecr,t)-\rho_\p(\vecr,t)}-\frac{2\chi}{\rho_\max}\rho_\p(\vecr,t)\biggr]\grad\rho_\p(\vecr,t)\nonumber\\&&+\frac{\rho_\p(\vecr,t)}{\rho_\max-\rho_\s(\vecr,t)-\rho_\ns(\vecr,t)-\rho_\p(\vecr,t)}\grad[\rho_\s(\vecr,t)+\rho_\ns(\vecr,t)]-\frac{2\chi\xi^2}{\rho_\max}\rho_\p(\vecr,t)\grad\nabla^2\rho_\p(\vecr,t).\label{eq:flux_p_p}
\end{eqnarray}
\end{widetext}
We keep the phosphorylation reaction kinetics the same as before, thereby assuming that a kinase roams freely everywhere in solution. However, the phosphorylation reactions require the encounter of a nonsticky protein and a phosphatase, which we model via mass action. Combining all of the above, the dynamics are given by
\begin{align}
    \frac{\partial\rho_\s(\vecr,t)}{\partial t}=&-\grad\cdot\vecJ_\s(\vecr,t)\nonumber\\&-k\rho_\s(\vecr,t)+\frac{k'}{\rho_\max}\rho_\p(\vecr,t)\rho_\ns(\vecr,t),\label{eq:PDEp_s}\\
    \frac{\partial\rho_\ns(\vecr,t)}{\partial t}=&-\grad\cdot\vecJ_\ns(\vecr,t)\nonumber\\&+k\rho_\s(\vecr,t)-\frac{k'}{\rho_\max}\rho_\p(\vecr,t)\rho_\ns(\vecr,t),\label{eq:PDEp_ns}\\
    \frac{\partial\rho_\p(\vecr,t)}{\partial t}=&-\grad\cdot\vecJ_\p(\vecr,t).\label{eq:PDEp_p}
\end{align}
In this version of the model, if there is transport, we observe propulsion of both condensates along the periodic $z$ axis. We scan various $k'$ and total concentration of phosphatase, and we find a propulsive band. In Fig.~\ref{fig:p-condensate}, we present a propulsive example with $\ell_k=40\xi$, $(\rho_\tot-\rho_\bin)/(\rho_\spn-\rho_\bin)=0.9$, $(\mu_\mol\kt/k')^{1/2}=\ell_k/5$, and $\rho_{\p,\tot}=1.075\rho_\bin$. 

\section{Supporting analytical calculations}\label{sec:surface}

In this appendix, we provide analytical calculations for the case without phase separation, supporting our use of the limit of instantaneous dephosphorylation at the colloid surface. We first compute the concentration profiles dictated by the phosphorylation cycles in the idealized instantaneous-dephosphorylation case. We then extend those expressions to two models for finite-rate dephosphorylation. None of these cases breaks rotational symmetry, so the colloid is immobile, $\vecv=\mathbf0$; thus, we only consider the radial component of the concentration profiles. 

The reason we devote a detailed appendix to the case of a finite dephosphorylation rate is that the formation of a condensate in our scheme requires an efficient phosphatase. It is therefore important to verify that a real phosphatase, operating at finite rates and over small spatial extent, is able to replenish sticky proteins at a sufficient rate to drive phase separation. 

\subsection{Instantaneous dephosphorylation}\label{sec:surface-id}

Prior to considering a phosphatase with a finite rate, we compute the concentration profile for instantaneous phosphorylation as in the main text. We ignore protein-colloid repulsion and discuss an ideal solution for simplicity (no phase separation). Equations~\eqref{eq:PDE_s} and~\eqref{eq:PDE_ns} at steady state simplify to
\begin{eqnarray}
    \frac1{r^2}\frac\d{\d r}[r^2J_{\s,r} (r)]&=&-k\rho_\s(r),\label{eq:PDE_s_id}\\
    \frac1{r^2}\frac\d{\d r}[r^2J_{\ns,r} (r)]&=&+k\rho_\s(r),\label{eq:PDE_ns_id}
\end{eqnarray}
with
\begin{equation}
    J_{i,r} (r)=-\mu_\mol\kt \frac{\d\rho_i(r)}{\d r},\qquad i=\s,\ns.\label{eq:flux_id}
\end{equation}
The boundary conditions here are instantaneous dephosphorylation [Eq.~\eqref{eq:BCinstantconv}], and there is no net protein influx into the colloid [Eq.~\eqref{eq:BCnoflux}], 
\begin{eqnarray}
    \rho_\ns(R)&=&0,\label{eq:BCinstantconv_id}\\
    J_{\s,r} (R)+J_{\ns,r} (R)&=&0.\label{eq:BCnoflux_id}
\end{eqnarray}
Furthermore, we enforce total saturation $\rho_\tot$ far from the droplet, and finite concentrations overall,
\begin{eqnarray}
    \rho_\s(\infty)+\rho_\ns(\infty)&=&\rho_\tot,\label{eq:BCconc_tot}\\
    \rho_\ns(\infty)&<&\infty.\label{eq:BCconc_finite}
\end{eqnarray}

The solution to Eqs.~\eqref{eq:PDE_s_id} and~\eqref{eq:PDE_ns_id} subject to the boundary conditions is straightforward, 
\begin{equation}
    \rho_\s(r)=\rho_\tot-\rho_\ns(r)=\rho_\tot\frac Rre^{-(r-R)/\ell_k},\label{eq:SSfield_id}
\end{equation}
where $\ell^2_k=\mu_\mol\kt/k$. Indeed, for $\rho_\tot>\rho_\bin$, there can potentially be a shell of sufficient material for forming droplets surrounding the colloid. The larger $\ell_k$, the more sticky material exists around the colloid, which is able to support progressively larger or more numerous droplets. Our aim with a fast phosphatase, therefore, is to simply have enough material to form at least one droplet around the colloid without requiring excessively large supersaturation $\rho_\tot$. We now investigate two models in which the phosphatase operates with a finite rate.

\subsection{Nonzero phosphatase volume}

Suppose the phosphatase occupies a particular ``catalytic volume,'' namely, a shell of width $w_\p$ around the colloid. This case modifies Eqs.~\eqref{eq:PDE_s_id} and~\eqref{eq:PDE_ns_id} to
\begin{eqnarray}
    \frac1{r^2}\frac\d{\d r}[r^2J_{\s,r} (r)]&=&-k\rho_\s(r)+k_\p\Theta(R+w_\p-r)\rho_\ns(r),\nonumber\\\label{eq:PDE_s_diffuse}\\
    \frac1{r^2}\frac\d{\d r}[r^2J_{\ns,r} (r)]&=&+k\rho_\s(r)-k_\p\Theta(R+w_\p-r)\rho_\ns(r),\nonumber\\\label{eq:PDE_ns_diffuse}
\end{eqnarray}
where $k_\p$ is the effective rate of the phosphatase, dependent on its areal density on the colloid surface, and $\Theta(x)$ is the step function, $\Theta(x>0)=1$ and $\Theta(x<0)=0$. As we explicitly included the phosphatase kinetics, we must enforce a flux-free boundary on the colloid surface for each protein species separately; the few nonsticky and majority sticky proteins that reach the surface must not flux through it. Thus, the previous boundary conditions [Eqs.~\eqref{eq:BCinstantconv_id} and~\eqref{eq:BCnoflux_id}] change to
\begin{equation}
    J_{i,r} (R)=0,\qquad i=\s,\ns.\label{eq:BCnoflux_diffuse}\\
\end{equation}

To solve Eqs.~\eqref{eq:PDE_s_diffuse} and~\eqref{eq:PDE_ns_diffuse}, we split the concentration profiles for $r<R+w_\p$ and $r>R+w_\p$, and enforce continuity of concentration and of flux at $r=R+w_\p$. Using the boundary conditions [Eqs.~\eqref{eq:BCconc_tot}, \eqref{eq:BCconc_finite}, and~\eqref{eq:BCnoflux_diffuse}], we find the general form
\begin{widetext}
\begin{equation}
    \rho_\s(r)=\rho_\tot-\rho_\ns(r)=\rho_\tot\times\begin{cases}
        \frac{k_\p}{k+k_{\p}}-A_<\frac{R+w_\p}{r}\left[\frac{R}{\ell_\p}\cosh\left(\frac{r-R}{\ell_\p}\right)+\sinh\left(\frac{r-R}{\ell_\p}\right)\right],
        & R<r<R+w_\p,\\
        A_>\frac{R+w_\p}re^{-[r-(R+w_\p)]/\ell_k},
        & r>R+w_\p,
    \end{cases}\label{eq:correct_target_blabla}
\end{equation}
where $\ell_\p^2=\mu_\col\kt/(k_\p+k)$ is the typical distance between surface conversions. Continuity of concentration and flux yields
\begin{eqnarray}
    1-\frac{\ell_\p^2}{\ell_k^2}-A_<\left[\frac{R}{\ell_\p}\cosh\left(\frac{w_\p}{\ell_\p}\right)+\sinh\left(\frac{w_\p}{\ell_\p}\right)\right]&=&A_>,\\
    A_<\left[\left(\frac{R}{\ell_\p}\frac{R+w_\p}{\ell_\p}-1\right)\sinh\left(\frac{w_\p}{\ell_\p}\right)+\frac{w_\p}{\ell_\p}\cosh\left(\frac{w_\p}{\ell_\p}\right)\right]&=&A_>\left(1+\frac{R+w_\p}{\ell_k}\right).\label{eq:SSfield_diffuse}
\end{eqnarray}
\end{widetext}

We are interested in the limit where $w_\p\ll R,\ell_k$ and $k_\p\gg k$ (so $\ell_\p\ll\ell_k$). To have a nonzero concentration of sticky proteins close to the sphere, we must operate in the regime $\ell_\p\ll R$, wherein the nonsticky-to-sticky conversions are able to counteract the diffusive outflux of sticky material down its concentration gradient. Independently of the magnitude of $w_\p/\ell_\p$, we keep only leading-order terms in $w_\p/R$, $\ell_\p/\ell_k$, and $\ell_\p/R$, and find 
\begin{equation}
    A_>=\frac1{1+(\ell_\p/R+\ell_\p/\ell_k)\coth(w_\p/\ell_p)}.
\end{equation}
Compare the $r>R+w_\p\simeq R$ case in Eq.~\eqref{eq:correct_target_blabla} to Eq.~\eqref{eq:SSfield_id}. Clearly, the finite reaction rate reduces the effective sticky concentration, $A_><1$. 

To approach the instantaneous conversion limit, where $A_>\to1$, we need at least $\ell^{-1}_\p\gg R^{-1}+\ell_k^{-1}$. If $w_\p\ll\ell_\p$, we need the stricter condition $\ell^{-2}_\p\gg w_\p^{-1}(R^{-1}+\ell_k^{-1})$. To determine how rapid the phosphatases must be, we express the dephosphorylation rate in terms of the phosphatase's catalytic efficiency $\eta_\p=k_{\mathrm{cat},\p}/K_{\mathrm{m},\p}$, surface concentration $\sigma_\p$, and catalytic width $w_\p$, $k_\p\simeq\eta_\p\sigma_\p w_\p^{-1}$. Thus, the condition on the phosphatase's catalytic efficiency is
\begin{equation}
    \eta_\p\gg \frac{\mu_\col\kt}{\sigma_\p}\left(\frac1R+\frac1{\ell_k}\right).\label{eq:CatEff_est}
\end{equation}
Taking the protein diffusion coefficient as $\mu_\mol\kt\sim100\micron^2/\s$, the distance $R,\ell_k\sim1\micron$, and a high surface density $\sigma_\p\sim1\mathrm{nm}^{-2}$, we obtain $\eta_\p\gtrsim10^{5}\mathrm{M}^{-1}\s^{-1}$.

To confirm that this model for a finite dephosphorylation rate does not alter the phenomena we outlined in the main text, we perform simulations of Eqs.~\eqref{eq:PDE_s} and~\eqref{eq:PDE_ns} with the additional reaction terms $\pm k_\p\Theta(R+w_\p-|\vecr|)\rho_\ns(\vecr)$ [Eqs.~\eqref{eq:PDE_s_diffuse} and~\eqref{eq:PDE_ns_diffuse}] and the altered boundary conditions [Eq.~\eqref{eq:BCnoflux_diffuse}]. The parameters we use are $w_\p=\xi$ and $\ell_\p=\ell_k/100$ and the even stricter case $\ell_\p=\xi$ (other parameters as in Appendix~\ref{sec:simulation-para}). Indeed, the phase diagram of Fig.~\ref{fig:diagram_static} simply shifts to higher $\ell_k$ and $\rho_\tot$ values.

\subsection{Linear adsorption theory}

Here, we ignore the volume occupied by phosphatases, and instead consider a model where nonsticky molecules should first adsorb to the colloid surface, undergo phosphorylation, and then desorb. Thus, the phosphatase does not operate anywhere in the bulk, and the steady concentration profiles follow Eqs.~\eqref{eq:PDE_s_id} and~\eqref{eq:PDE_ns_id}. However, the boundary conditions become more complicated as follows: Instead of Eqs.~\eqref{eq:BCinstantconv_id} and~\eqref{eq:BCnoflux_id}, the flux of each protein type at the boundary is given in terms of adsorption and desorption rates,
\begin{equation}
    J_{i,r}(R)=r_{i,\d}\sigma_i-r_{i,\a}\rho_i(R),\qquad i=\s,\ns,\label{eq:BCflux_ads_def}
\end{equation}
where $\sigma_i$ is the surface density of adsorbed $i$-type protein (note that we consider steady state in a problem that does not break spherical symmetry), and $r_{i,\a}$ and $r_{i,\d}$ are adsorption and desorption rates, respectively. 

Now, it is the adsorbed nonsticky proteins that are being converted unto sticky proteins with a rate $r_\p$. Assuming that the enzymes (binding sites) do not approach saturation, we write the steady-state conditions
\begin{eqnarray}
    0&=&r_{\s,\a}\rho_\s(R)-r_{\s,\d}\sigma_\s+r_\p\sigma_\ns,\\
    0&=&r_{\ns,\a}\rho_\ns(R)-r_{\ns,\d}\sigma_\ns-r_\p\sigma_\ns.
\end{eqnarray}
Solving for $\sigma_\ns$, we find the closed form for the boundary conditions, Eq.~\eqref{eq:BCflux_ads_def}:
\begin{equation}
    J_{\s,r}(R)=-J_{\ns,r}(R)=\kappa_\p\rho_\ns(R),\label{eq:BCflux_ads}
\end{equation}
where $\kappa_\p=r_{\ns,\a}/(1+r_{\ns,\d}/r_\p)$ is the net reaction rate, which closely relates to the Michaelis-Menten reaction rate.

Solving Eqs.~\eqref{eq:PDE_s_id} and~\eqref{eq:PDE_ns_id} with the boundary conditions Eqs.~\eqref{eq:BCconc_tot}, \eqref{eq:BCconc_finite}, and~\eqref{eq:BCflux_ads}, we obtain
\begin{equation}
    \rho_\s(R)=\rho_\tot-\rho_\ns(R)=\frac{\rho_\tot}{1+(\ell_\p/R+\ell_\p/\ell_k)}\frac Rre^{-(r-R)/\ell_k}.\label{eq:SSfield_ads}
\end{equation}
where $\ell_\p=\mu_\mol\kt/\kappa_\p$. Compare Eq.~\eqref{eq:SSfield_ads} to Eq.~\eqref{eq:SSfield_id}. Once again, a finite reaction rate reduces the effective sticky-protein concentration. 

As before, to approach the instantaneous conversion limit, we need at least $\ell^{-1}_\p\gg R^{-1}+\ell_k^{-1}$, which restricts the dephosphorylation rate as follows. In term of the phosphatase's catalytic efficiency $\eta_\p\equiv k_{\mathrm{cat},\p}/K_{\mathrm{m},\p}$ and surface concentration $\sigma_\p$, the net surface reaction rate is $\kappa_\p\simeq\eta_\p\sigma_\p$. With this rate in hand, the condition on the phosphatase catalytic efficiency coincides exactly with Eq.~\eqref{eq:CatEff_est}.
Thus, with the above parameter values, we once again arrive at the requirement $\eta_\p>10^{5}\mathrm{M}^{-1}\s^{-1}$.

To confirm that this model for finite dephosphorylation rate does not alter the phenomena we outlined in the main text, we perform simulations of Eqs.~\eqref{eq:PDE_s} and~\eqref{eq:PDE_ns} with the altered Robin boundary conditions (Eq.~\eqref{eq:BCflux_ads}), where $\ell_\p=\xi$. Once again, the phase diagram of Fig.~\ref{fig:diagram_static} is simply shifted to higher $\ell_k$ and $\rho_\tot$ values.

\subsection{Estimating energetic efficiency}\label{sec:surface-cat}

In this section, we detail the estimates behind the energy consumption stated for our system in Sec.~\ref{sec:discussion-thermo}. Consider the $R=0.2\micron$ colloid. Its Stokes mobility is $\mu_\col=(6\pi\eta R)^{-1}\sim300\micron/(\mathrm{pN}\cdot\s)$, and when mobile it experiences forces of order $F\sim 0.3\mathrm{pN}$ (Fig.~\ref{fig:diagram_free}). Therefore, the energy investment needed to oppose the friction experienced by the colloid is $\mu_\col F^2\sim10^4k_\mathrm{B}T/\s$. 

Next, we estimate how many ATPs are consumed by the kinase. Ignoring the condensate, the profile of sticky molecules is approximately Eq.~\eqref{eq:SSfield_id}. The total number of reactions per second is about $\int_R^\infty [k\rho_\s(r)]4\pi r^2\d r=4\pi\varrho_\mathrm{tot} DR(1+R/\ell_k)\sim10^{10}\s^{-1}$. Assuming each reaction consumes one ATP, each storing about $10k_\mathrm{B}T$~\cite{MiloBOOK2015}, the total energy consumed is about $10^{11}k_\mathrm{B}T/\s$.

\end{document}